\documentclass[amsmath,twocolumn,superscriptaddress,nature,aps]{revtex4-2}
\usepackage{amsthm,amsfonts,graphicx,color,amsmath,amssymb,textcomp}
\usepackage{graphicx}
\usepackage{wasysym}
\usepackage{amsfonts}
\usepackage{bm}
\usepackage{enumerate}
\usepackage{color}
\usepackage{epstopdf}
\usepackage{latexsym}
\usepackage[breaklinks,colorlinks = true,linkcolor = red,urlcolor=cyan,citecolor=red]{hyperref}
\usepackage[caption=false,singlelinecheck=false]{subfig}

\usepackage[compat=1.1.0]{tikz-feynman}
\usepackage{times}
\usepackage{float}
\usepackage{changes}
\colorlet{Changes@Color}{black}

\usepackage{times}
\newcommand{\bea}{\begin{eqnarray}}
\newcommand{\eea}{\end{eqnarray}}
\newcommand{\be}{\begin{eqnarray}}
\newcommand{\ee}{\end{eqnarray}}
\newcommand{\bw}{\begin{widetext}}
\newcommand{\ew}{\end{widetext}}
\newcommand{\nn}{\nonumber}

\newcommand{\la}{\langle}
\newcommand{\ra}{\rangle}

\newcommand{\tbf}{\textbf}

\usepackage{filecontents}
\usepackage{multibib}
\newcites{supp}{Supplementary References}

\begin{document}
\title{Optical absorption signatures of superconductors driven by Van Hove singularities}

\author{Hyeok-Jun Yang}
\email{hyang23@nd.edu}
\affiliation{Department of Physics, University of Notre Dame, Notre Dame, Indiana 46556, USA}

\author{Yi-Ting Hsu}
\email{yhsu2@nd.edu}
\affiliation{Department of Physics, University of Notre Dame, Notre Dame, Indiana 46556, USA}

\date{\today}

\begin{abstract}
Due to the diverging density of states (DOS), Van Hove singularities (VHS) near the Fermi level are known to boost the susceptibility to a wide variety of electronic instabilities, including superconductivity. 
We theoretically show that the number of VHS in the normal state can be qualitatively inferred from the optical absorption spectra Re $\sigma_{ii}(\omega)$ in the superconducting state. 
The key feature is the absorption peak at frequency $\omega=2\Delta$ from the optical transition across the superconducting gap $\Delta$, which is forbidden in a single-band clean superconductor when the inversion symmetry is preserved and the band is quadratic.  
Although VHS dispersions are mostly quadratic, we find that a divergent peak occurs when there are multiple VHS on the Fermi surface under an applied current. In contrast, we find non-diverging weak peaks when there is only a single VHS. Depending on whether this single VHS has logarithmically or power-law divergent DOS, the peaks in Re $\sigma_{xx}(\omega)$ and Re $\sigma_{yy}(\omega)$ are nearly isotropic and anisotropic, respectively. 
Therefore, we propose that experimentally measured peak magnitude and anisotropy in the optical absorption spectra of VH-driven superconductors can be used to determine the number and type of VHS on the Fermi surface.
Disorders could even facilitate in distinguishing the multiple and single VHS scenarios since only the diverging peak in the latter case is expected to survive.

\end{abstract}
\maketitle
\label{sec:Introduction}
\textit {Introduction ---} Van Hove singularities (VHS) are saddle points in electronic structures \cite{PhysRev.89.1189, lifshitz1960anomalies}, which  can often be found near the Fermi level in Van der Waals materials, such as Kagome metals \cite{PhysRevMaterials.3.094407, PhysRevMaterials.5.L111801, Jiang2021, Zhao2021,Duan2021, Xiang2021, PhysRevX.11.031050, PhysRevX.11.041030, PhysRevLett.127.046401, Neupert2022, 10.1093/nsr/nwac199, Hu2022, Scammell2023}, various Moire superlattices \cite{doi:10.1021/acsnano.0c10435, doi:10.1073/pnas.1108174108,doi:10.1021/acs.nanolett.6b01906,PhysRevLett.109.126801, Schouteden2016, PhysRevLett.109.196802, PhysRevB.99.195120, PhysRevB.100.085136}, and graphene-based few-layers heterostructure \cite{Li2010, PhysRevLett.106.126802, PhysRevB.90.115402,Liao2015,Yin2016, Cao2018, Cao2018_2, Kerelsky2019-rq, Xie2019, doi:10.1126/science.aav1910, Nimbalkar2020, Xu2021, doi:10.1126/science.abm8386, Chichinadze2022, Zhang2023}. 
Due to the diverging density of states (DOS) at VHS, the electronic correlations among these VH ‘hot spots’ on the Fermi surface can drive a rich variety of symmetry-broken and topological phases \cite{Nandkishore2012, PhysRevLett.108.227204, Hsu_graphene,PhysRevX.11.021024,PhysRevB.98.214521, PhysRevB.102.125120, Hsu_TMD,PhysRevB.104.035142}, including conventional and unconventional superconductivity \cite{HUR20091452, Nandkishore2012, PhysRevB.86.115426, Hsu_graphene,PhysRevB.102.125141,Hsu_TMD,PhysRevB.106.174514, PhysRevB.95.035137, PhysRevB.99.144507, PhysRevX.8.041041}. 
Instead of detailed Fermi surface shapes, the pairing symmetry is predominantly determined by    
the number of conventional VHS (cVHS) \cite{HUR20091452, Nandkishore2012, PhysRevB.104.045122} and higher-order VHS (hVHS) \cite{PhysRevLett.109.176404, PhysRevB.102.245122, PhysRevLett.123.207202, Yuan2019,PhysRevResearch.1.033206, PhysRevB.102.125141, PhysRevB.105.235145}, where the former and latter are characterized by logarithmically and power-law divergent DOS, respectively. 
It is therefore curious whether the VHS type in the normal state can be inferred from the experimental observables in the superconducting state. 

Optical measurements are widely utilized experimental techniques for investigating electronic structures, broken symmetries, and quantum geometric properties across diverse quantum phases. 
In superconductors, the superfluid weight extracted from the dissipative optical conductivity Re $\sigma(\omega\rightarrow 0)$  \cite{Peotta2015, PhysRevLett.117.045303, PhysRevB.95.024515, PhysRevB.96.064511, ROSSI2021100952, PhysRevB.104.L100501} as well as the angular-resolved photoemission spectroscopy \cite{PhysRevLett.126.187001, liao2023unveiling} 
were both theoretically proposed to reflect the quantum geometry of quasiparticles.  
Experimentally, the gap magnitude $|\Delta|$ has also been determined from finite-frequency optical conductivity Re $\sigma(\omega)$ in disordered superconductors \cite{PhysRevB.25.1565, PhysRevLett.76.1525, PhysRevB.93.180511,PhysRevB.96.144507,  Uzawa2020}. 
In a clean superconductor, an optical absorption peak at a frequency $\omega=2\Delta$ \cite{PhysRevLett.122.257001, PhysRevLett.125.097004} was observed only in the presence of an applied dc supercurrent \cite{PhysRevLett.122.257001, PhysRevLett.125.097004}. The $2\Delta$ absorption peak was understood to result from the transition across the superconducting gap $\Delta$ (see Fig. 1a).   

This optical transition across the gap in a clean single-band superconductor is theoretically known to be activated only when (1) the inversion symmetry is broken, and (2) the current $\textbf{j}$ is not conserved. Selection rule (1) has been well-studied, and can be controllably broken by applying a dc supercurrent \cite{PhysRevB.100.220501, Ahn2021,PhysRevB.106.214526, PhysRevB.106.L220504}. Selection rule (2) can be understood as follows \cite{https://doi.org/10.1002/andp.200651807-809, SM_optical}:  
The transition amplitude is proportional to the matrix element $\langle \Phi_{\text{excited}}|\textbf{j}|\Phi_{\text{GS}}\rangle$, where $|\Phi_{\text{GS}}\rangle$ and $|\Phi_{\text{excited}}\rangle$ are the superconducting ground and excited states. This matrix element strictly vanishes when the eigenstates preserve the current $\textbf{j}$, which happens in the limit of parabolic normal bands. 
When lattice effects come into play, such as non-parabolic band structures, different degrees of current relaxation can naturally occur. 
However, since bands are often approximately quadratic near the bottom,
an evident $2\Delta$ absorption peak is still not generally expected in clean superconductors.

In this work, we show that superconductivity driven by VHS can exhibit a prominant  optical absorption peak at $\omega=2\Delta$ under an applied supercurrent, even when the pairing gap is $s$-wave. Importantly, we find that the divergence and anisotropy of the peak are qualitatively determined by the number and type of VHS in the normal state. 
specifically, the cucally, by calculating the dissipative optical conductivity Re $\sigma_{ii}(\omega)$ in the presence of VH-induced current relaxation, we find that the peak is logarithmically divergent when there are multiple cVHS near the Fermi surface. In contrast, the peak is non-diverging for normal states with only one cVHS or hVHS. We propose to experimentally distinguish the single cVHS and hVHS cases by the anisotropy between $\sigma_{xx}(\omega)$ and $\sigma_{yy}(\omega)$ (see Fig. 1a).


\label{sec:Non-interacting dispersion}
\textit{Conventional and higher-order VHS ---} 
For a VHS located at momentum $\tbf{M}$, we consider the following non-interacting dispersion 
\bea
&&
\epsilon_{\tbf{q}}=\Big(\frac{\alpha_1}{2}q_x^2+
\frac{\alpha_2}{12}q_x^4\Big)
-\Big(
\frac{\beta_1}{2}q_y^2+
\frac{\beta_2}{12}q_y^4
\Big)+O(q^6), 
\label{eq:VHS}
\eea
where $\tbf{q}=\tbf{k}-\tbf{M}$ is the crystal momentum measured from $\tbf{M}$. Here, 
$\epsilon_{\tbf{q}}$ is valid within a patch centered at $\tbf{M}$ with a width $\Lambda_{i}$ in $i=x,y$ directions. 
This dispersion describes a cVHS or hVHS under the following conditions  
\bea
\text{cVHS}: 
&~~&
\alpha_1 \gg |\alpha_2|,|\beta_2|
,\quad \beta_1 \gg |\alpha_2|,|\beta_2|,
\nn\\
\text{hVHS}:
&~~& 
\alpha_1\gg |\alpha_2|, \quad \beta_1=0, \quad \beta_2>0,  
\label{eq:VHS_type}
\eea
where we assume the lattice constant to be unity.
For a cVHS, the dispersion is quadratic in both $q_x$ and $q_y$ so that the DOS diverges logarithmically. 
For a hVHS, the dispersion becomes quartic in $q_y$ while remaining quadratic in $q_x$. The resulting DOS thus exhibits a stronger power-law divergence.



Besides the divergence in DOS,
another key difference between cVHS and hVHS is the effective mass. Close to the saddle point $\tbf{M}$, the effective mass of cVHS is a momentum $\textbf{q}$-independent tensor, $\mathbb{M}^{-1}_{\tbf{q}}=\text{diag}(\alpha_1,-\beta_1)$ while that of the hVHS is $\textbf{q}$-dependent, $\mathbb{M}^{-1}_{\tbf{q}}=\text{diag}(\alpha_1,-\beta_2q_y^2)$. 
We will show in the following that this difference in the effective mass leads to qualitative differences in the optical conductivity for superconductivity driven by cVHS and hVHS.  

\begin{figure}[t!]
{\includegraphics[width=0.48\textwidth]{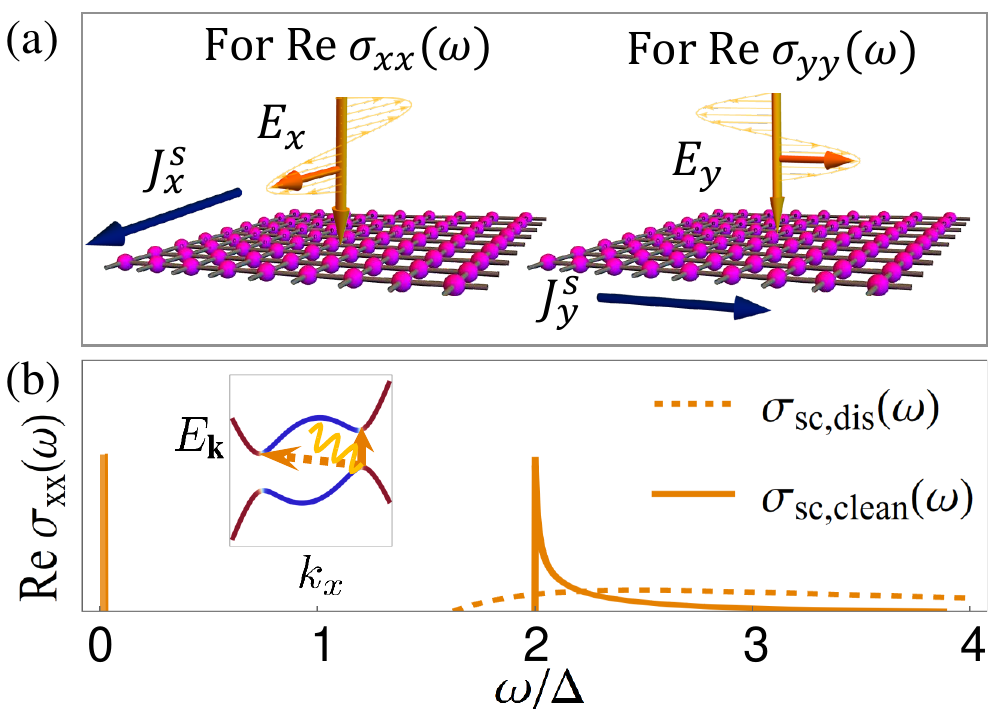}}
\caption{
(a) Schematic experimental setups for the measurements of Re $\sigma_{xx}$ and Re $\sigma_{yy}$. The pink layer and orange arrow represent a superconducting film under a linearly polarized light, where $E_i(t)=E_0e^{-i\omega t}$ is the corresponding electric field in $i=x,y$ directions. A supercurrent $J^s_i$, which shifts the Fermi surface by a momentum $Q_i/2$ is applied along the same direction $i=x,y$ to break the inversion symmetry. (b) Schematic illustration of the optical absorption spectra expected from a clean superconducting state
(solid line) and disorders (dashed line). The solid peak at frequency $\omega=2\Delta$ results from the momentum-preserving transition across the gap, whereas the disorder-driven response results from the indirect transitions (see the inset). The zero-frequency solid peak results from the superfluid weight. 
}
\label{fig:schematic}
\end{figure}

\label{sec:Optical conductivity in superconducting states}
\textit {Model ---}
To allow non-vanishing optical absorption $\text{Re}\sigma_{ii}(\omega)$ in the superconducting state, we propose to break the inversion symmetry by applying a supercurrent and achieve current relaxation by tuning the chemical potential $\mu$ to VH filling.  
The BdG Hamiltonian that describes such a VH-driven superconductor is given by
\bea
H^{\text{BdG}}_{\tbf{q}}(\tbf{Q})
=
\begin{pmatrix}
\xi_{\tbf{q}+\frac{\tbf{Q}}{2}} & \Delta \\
\Delta^{*} & -\xi_{-\tbf{q}+\frac{\tbf{Q}}{2}}
\end{pmatrix}, 
\label{eq:BdG_Hamiltonian}
\eea
written in the Nambu basis $\Psi_{\tbf{q}}=(c_{\tbf{q}+\frac{\tbf{Q}}{2}\uparrow}, c_{-\tbf{q}+\frac{\tbf{Q}}{2}\downarrow}^\dagger)^T$. 
Here, the normal-state dispersion $\xi_{\tbf{q}}$ is shifted by $\tbf{Q}/2$ in momentum due to the applied current, and $c^{\dagger}_{\textbf{q}s}$ creates an electron with spin $s=\uparrow,\downarrow$ at momentum
$\tbf{q}$ away from the VH point.
The normal-state dispersion  $\xi_{\tbf{q}}=\epsilon_{\tbf{q}}-\mu$ exhibits one or more cVHS or hVHS when the chemical potential $\mu=0$ (see Eq. \ref{eq:VHS}). We assume the superconducting order parameter $\Delta$ is momentum-independent, which can be determined by self-consistenly solving the gap equation in the presence of an on-site interaction $H^{\text{int}}
=U\sum_{\tbf{k}\tbf{k}'\tbf{p}\sigma\sigma'}c^{\dagger}_{\tbf{k}+\tbf{p}\sigma}c^{\dagger}_{\tbf{k}'-\tbf{p}\sigma'}c_{\tbf{k}'\sigma'}c_{\tbf{k}\sigma}$ with $U<0$ (see supplementary material (SM) \cite{SM_optical} Sec. \ref{sec:SM_GI}). 
Since the low-energy dispersions $\xi$ are not isotropically parabolic near the saddle point(s), selection rule (2) is fulfilled for the optical absorption in the superconducting state described by 
$H^{\text{BdG}}_{\tbf{q}}(\tbf{Q})$ 
. 

Within this formalism, the total current density is given by $\tbf{J}=\frac{1}{V}\sum_{\tbf{q}}\Psi_\tbf{q}^\dagger {\tbf{j}}_\tbf{q}\Psi_\tbf{q}$,  where $V$ is the volume and the current operator has the form 
\bea
\tbf{j}_\tbf{q}={\tbf{v}}_\tbf{q}\tau^0 + \mathbb{M}_{\tbf{q}}^{-1}\cdot \frac{\tbf{Q}}{2}~\tau^3  
\label{eq:bare_current_BdG}
\eea
up to the lowest order in $|\tbf{Q}|$, and $\tau_i$ denotes Nambu-basis Pauli matrices. Importantly, the second term breaks the inversion symmetry so that selection rule (i) is fulfilled. Thus, the resulting absorption amplitude is expected to be proportional to the effective mass $(\mathbb{M}_{\tbf{q}}^{-1})_{ij}=\partial_{q_i}\partial_{q_j}\xi_\tbf{q}$ in the normal state. 

\label{sec:Gauge-invariant response}
\textit{Optical conductivity with vertex correction---} The dissipative optical conductivity in the superconducting state can be calculated by the standard linear response theory as \cite{bruus2004many}
\bea
\text{Re}~\sigma_{ii}(\omega)=\text{Im}\frac{P_{ii}(i\omega_m \rightarrow\omega +i\eta)}{\omega},
\label{eq:def_optical}
\eea 
where $P_{ii}(i\omega_m)$ is the current-current correlator and $\omega_m$ is the bosonic Mastubara frequency. 
Importantly, it was shown that the current relaxation due to lattice effects cannot be correctly captured if $P_{ii}(i \omega_m)$ is calculated from the BdG Hamiltonian $H^{\text{BdG}}_{\textbf{q}}$ since the Ward identity \footnote{The Ward identity has the form $p_\mu \tilde{\Gamma}_{\tbf{q}+\tbf{p},\tbf{q}}^{\mu}(iq_n +i\omega_m,iq_n)=G^{-1}_{iq_n +i\omega_m,\tbf{q}+\tbf{p}}\tau^3 
-\tau^3 G^{-1}_{iq_n,\tbf{q}}$ where $\Gamma_{\tbf{q},j}(i\omega_m)= \tilde{\Gamma}^{j}_{\tbf{q},\tbf{q}}(iq_n+i\omega_m,iq_n)$.} is violated under the mean-field approximation. 
To restore the Ward identity, it has been shown that it is sufficient to include the corrections from the pairing interaction in $P_{ii}(i\omega_m)$ under the random phase approximation (RPA) \cite{schrieffer2018theory, PhysRevB.95.014506}. Specifically, the current-current correlator under RPA is given by 
\bea
&&P_{ij}(i\omega_m)=
\nn\\
&&~~
-T\sum_{iq_n}\frac{1}{V}\int_\tbf{q}\text{Tr}
\Big[
j_{\tbf{q},i} G_{iq_n +i\omega_m,\tbf{q}}
\Gamma_{\tbf{q},j}(i\omega_m) 
G_{iq_n,\tbf{q}}\Big],\quad
\label{eq:correlator}
\eea
where $T$ is the temperature, $q_n$ is the fermionic Mastubara frequency, and $G_{iq_n,\tbf{q}}=(iq_n -H^{\text{BdG}}_{\tbf{q}})^{-1}$ is the Green function obtained from the BdG Hamiltonian. 
Here, $j_{\tbf{q},i}$ and $\Gamma_{\tbf{q},j}(i\omega_m)=j_{\tbf{q},j}+\delta \Gamma_{\tbf{q},j}(i\omega_m) ~ (i=x,y)$  are the bare and dressed current operators, where the latter can be obtained by self-consistently solving the following Bethe–Salpeter equation, 
\begin{figure}[t!]
{\includegraphics[width=0.48\textwidth]{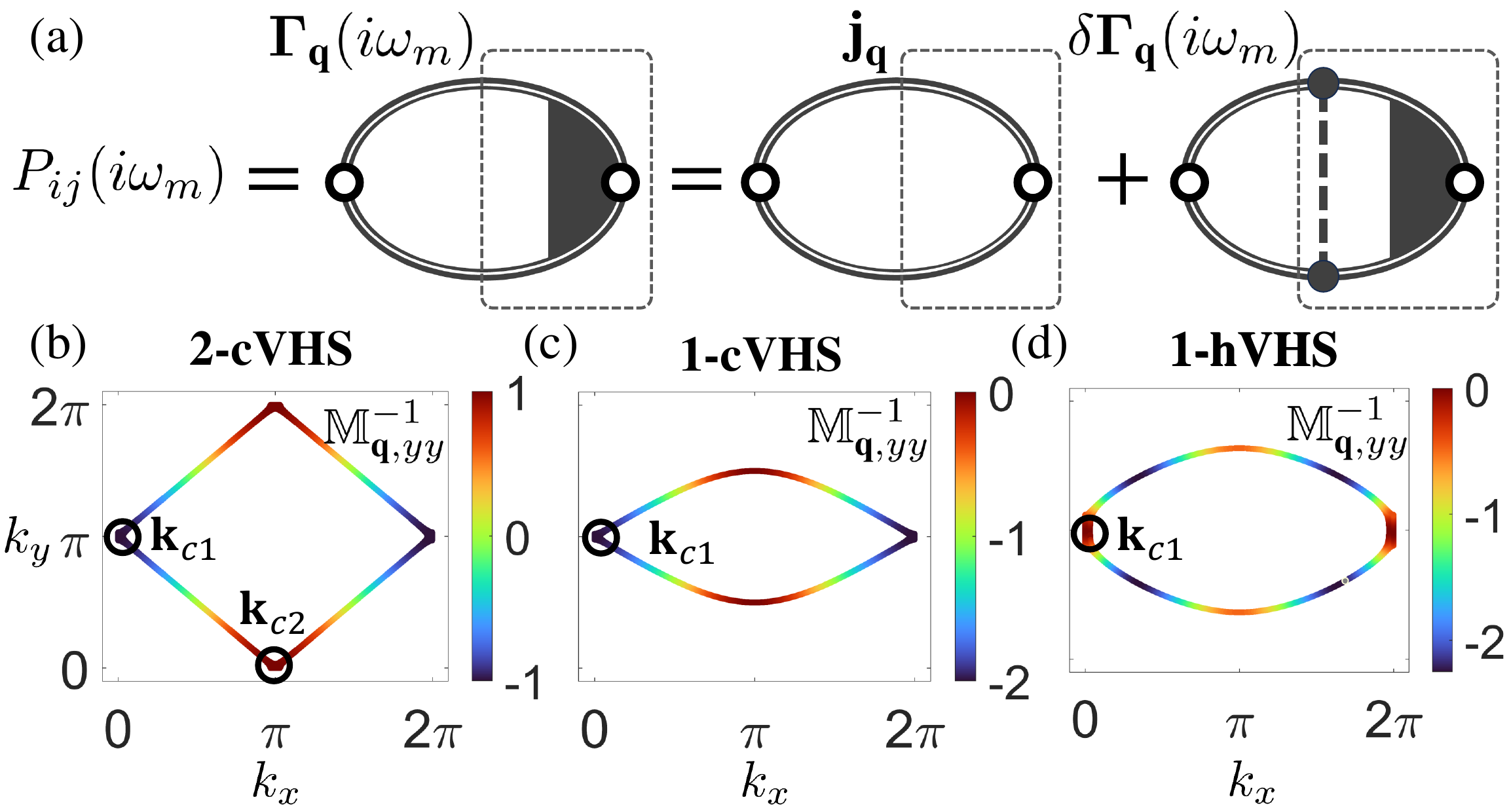}}
\caption{
(a) Diagrammatic representations for the total current-current correlator in Eq. \ref{eq:correlator}. 
The open circle, filled circle, double line, dotted line, and shaded region represent the current vertex, interaction vertex, dressed Green function, density-density interaction, and dressed current operator in Eq. \ref{eq:vertex_operator}, respectively. (b)-(d) The inverse effective mass $\mathbb{M}_{\tbf{q},yy}^{-1}$ along the Fermi surface for the lattice models used to calculate Fig. \ref{fig:plot} (a)-(c). These three models show (b) two cVHS at momenta $\tbf{k}_{c1}$ and $\tbf{k}_{c2}$, (c) a single cVHS at $\tbf{k}_{c1}$, and (d) a single hVHS at $\tbf{k}_{c1}$ on the Fermi surface.}
\label{fig:diagram}
\end{figure}
\bea
&&\boldsymbol{\Gamma}_{\tbf{q}}(i\omega_m)
=\tbf{j}_\tbf{q}
\nn\\
&&~~
-T\sum_{iq_n}\frac{1}{V}\int_{\tbf{q}'}U\tau^3 G_{iq_n+i\omega_m,\tbf{q}'} \boldsymbol{\Gamma}_{\tbf{q}'}(i\omega_m)
G_{iq_n,\tbf{q}'}\tau^3,  ~~~
\label{eq:vertex_operator}
\eea
with $\tau_i$ being the Nambu Pauli matrices. The corresponding ladder diagrams for the resulting current-current correlator $P_{ii}(i\omega_n)$ are shown in Fig. \ref{fig:diagram}. 
Finally, the resulting optical conductivity obtained from Eqs. \ref{eq:def_optical}-\ref{eq:vertex_operator} is given by
\bea
&&
\text{Re}~\sigma_{ii}(\omega)
=\sigma^{(0)}_{ii}(\omega)
+\delta\sigma_{ii}(\omega)
\nn\\
&&
~~=\Theta(\omega-2|\Delta|)\frac{\pi Q_i^2|\Delta|^2}{2\omega^2\sqrt{ \omega^2-4|\Delta|^2}}(\mathcal{C}_{ii}^{(0)}+\delta \mathcal{C}_{ii}),~~
\label{eq:bare+vertex}
\eea
where $\sigma^{(0)}_{ii}$ is the bare optical conductivity and $\delta\sigma_{ii}$ is the correction resulting from the vertex correction $\delta\Gamma_{\textbf{q}}$ to the current (see SM \cite{SM_optical} Sec. \ref{sec:SM_Optical}).

The frequency dependence of the optical conductivity $\text{Re}~\sigma_{ii}(\omega)$ in Eq. \ref{eq:bare+vertex} 
describes a peak at $\omega=2|\Delta|$, which results from the transition across the superconducting gap. For all the numerical calculations in this work, we regulate the peak by assuming a fixed finite scattering rate $\Gamma >0$ \cite{foot1} so that $\frac{1}{\sqrt{(\omega/2)^2-|\Delta|^2}}\rightarrow \frac{1}{\sqrt{|\Delta|\Gamma}}$. This is generally expected from reasons such as disorders.
The intensity of this peak is determined by the order parameter $\Delta$, the current momentum $Q_i$, as well as the normal state properties captured in the intensity factors $\mathcal{C}^{(0)}_{ii}$ and $\delta \mathcal{C}_{ii}$. 
Specifically, these factors are determined by the effective mass $\mathbb{M}_{\tbf{q},ii}$ and the DOS as   
\bea
&&
\mathcal{C}_{ii}^{(0)}=\int_{\tbf{q}}
\delta(\xi_\tbf{q})(\mathbb{M}_{\tbf{q},ii}^{-1})^2,
\;\;
\delta \mathcal{C}_{ii}=-\frac{\Big(\int_{\tbf{q}}
\delta(\xi_\tbf{q})\mathbb{M}_{\tbf{q},ii}^{-1}\Big)^2}{\int_\tbf{q}\delta(\xi_\tbf{q})},\;\;\quad
\label{eq:intensity}
\eea
where $\mathcal{C}^{(0)}_{ii}$ and $\delta \mathcal{C}_{ii}$ originate from the bare current operator $\tbf{j}_\tbf{q}$ and the vertex correction $\delta\boldsymbol{\Gamma}_{\tbf{q}}(iq_n)$ in Eq. \ref{eq:vertex_operator}, respectively (see SM \cite{SM_optical} Sec. \ref{sec:SM_Optical}).

Note that it is clear from Eq. \ref{eq:bare+vertex} and \ref{eq:intensity} how selection rules (i) and (ii) forbid  the optical absorption. For selection rule (i), Re $\sigma_{ii}(\omega)=0$ when the inversion symmetry is restored by removing the applied current, i.e. $Q_i=0$. 
For selection rule (ii), 
the current is conserved when the effective mass $\mathbb{M}_{\tbf{q},ii}$ is momentum $\tbf{q}$-independent. 
In this case, Eq. \ref{eq:intensity} indicates that the intensity factors perfectly cancel each other $\mathcal{C}_{ii}^{(0)}+\delta \mathcal{C}_{ii}=0$ so that Re $\sigma_{ii}(\omega)=0$. 
Since both intensity factors carry divergence from the VHS density of states, when selection rule (ii) is violated, the net intensity $\mathcal{C}_{ii}^{(0)}+\delta \mathcal{C}_{ii}$ can be diverging or non-diverging depending on the cause and degree of current relaxation. In the following, we examine how the net intensity and its anisotropy between $i=x,y$ depend on the number and type of VHS near the Fermi surface.  




\textit{Superconductivity driven by cVHS---} 
First, we investigate the optical absorption Re $\sigma(\omega)$ in a superconducting state where the Fermi surface contains multiple or one cVHS. 
Multiple cVHS can naturally occur in the presence of crystalline symmetries, such as discrete rotational symmetries $\mathcal{R}_n$, where there are $n_{\text{VH}}=n/2$ and $n$ cVHS located at momenta $\tbf{M}$, $\mathcal{R}_n\tbf{M},\cdots$, for even and odd $n$, respectively, at $\xi_\tbf{q}=0$. 
The case of a single cVHS can then be achieved by breaking the $\mathcal{R}_n$ symmetry with external perturbations. For instance, there are typically two $\mathcal{R}_4$-related VH points $\textbf{k}=(\pi,0)$ and $(0,\pi)$ on a square lattice. Under an uniaxial strain that breaks the $\mathcal{R}_4$ into a $\mathcal{R}_2$ symmetry, generally only one of the cVHS survives on the Fermi surface.  

For both $n_{\text{VH}}>1$ and $n_{\text{VH}}=1$ cVHS cases, the bare intensity $\mathcal{C}_{ii}^{(0)}$ of the absorption peak at $\omega=2\Delta$ (see Eq. \ref{eq:bare+vertex}) has the general form 

\bea
&&
\mathcal{C}_{\text{cVHS},xx}^{(0)}
=
a^{(0)}_{xx}
\log(\frac{\Lambda_x}{\sqrt{\mu/(2\alpha_1) }})+f_{xx}^{(0)},
\nn\\
&&
\mathcal{C}_{\text{cVHS},yy}^{(0)}= 
a^{(0)}_{yy}
\log(\frac{\Lambda_y}{\sqrt{\mu/(2\beta_1) }})
+f_{yy}^{(0)}, 
\;\;\quad
\label{eq:bare_cVHS}
\eea 
which contains a log-divergent term resulting from the DOS of cVHS. The coefficients $a^{(0)}_{xx (yy)}$ and the non-divergent terms $f_{xx (yy)}^{(0)}$ depend on
the patch widths $\Lambda_{x(y)}$ and dispersion coefficients $\alpha_{1,2} (\beta_{1,2})$ in Eq. \ref{eq:VHS}. See SM Sec. \ref{sec:SM_Peak_intensity} \cite{SM_optical} for the explicit forms in orthorhombic $n=2,4$ and hexagonal $n=3$ systems.  

The vertex correction $\delta \mathcal{C}_{ii}$, on the other hand, is sensitive to the number $n_{\text{VH}}$ of cVHS. 
\begin{itemize}
    \item Multiple cVHS $n_{\text{VH}}>1$ 
\end{itemize}

\begin{figure}[t!]
{\label{fig:plot}
\includegraphics[width=0.48\textwidth]{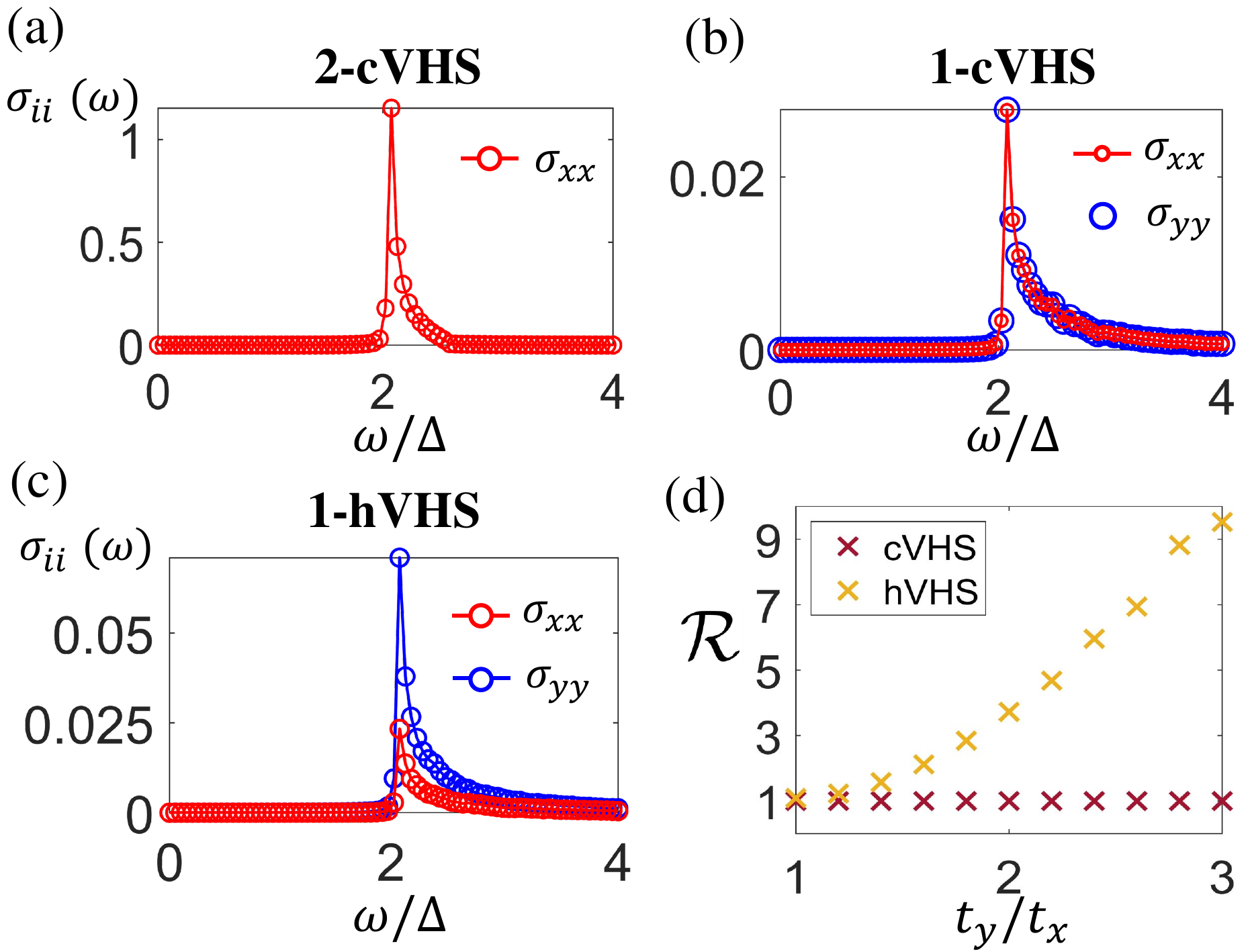}}
\caption{
Optical conductivity numerically calculated for clean superconducting lattice models (defined in the text) with different VH patterns on the Fermi surface: 
(a) two cVHS, (b) single cVHS, (c) single hVHS. All energy scales are expressed in unit of hopping parameters $t=t_x=1$. The rest of the hopping parameters are set to be $t_y=2,t_{yy}=0$ in (b), and $t_y=2, t_{yy}=0.5$ in (c). The sizes of superconducting gap and supercurrent momentum are $|\Delta|=0.1$, $|\tbf{Q}|=0.02$ respectively.
(d) The ratio $\mathcal{R}\equiv \text{Re}~ \sigma_{yy}/\text{Re}~\sigma_{xx}$ as a function of $t_y/t_x$ for the single cVHS and hVHS.}
\label{fig:plot}
\end{figure} 
When there are $n_{\text{VH}}>1$ cVHS near the Fermi level, we calculate Eq. \ref{eq:intensity} and find the vertex corrections to be  
\bea
\delta \mathcal{C}_{\text{cVHS},xx} 
&=&
\delta a_{xx}
\log(\frac{\Lambda_x}{\sqrt{\mu/(2\alpha_1) }})
+
\delta f_{xx},
\nn\\
\delta \mathcal{C}_{\text{cVHS},yy} 
&=&
\delta a_{yy}
\log(\frac{\Lambda_y}{\sqrt{\mu/(2\beta_1) }})
+
\delta f_{yy},
,
\label{eq:corr_multiple}
\eea
where $\delta a_{xx(yy)}<0$ and $\delta f_{xx(yy)}<0$ are also analytic functions of $\Lambda_{x(y)}$ and $\alpha_{1 (2)}, \beta_{1 (2)}$.
Similar to the bare contributions $\mathcal{C}^{(0)}_{\text{cVHS},ii}$, the vertex corrections $\delta \mathcal{C}_{\text{cVHS},ii}$ contain a log divergent part that comes from the density of states and a non-divergent part that depends on the cVHS dispersions. 
Importantly, the log divergent terms in $\mathcal{C}^{(0)}_{\text{cVHS},ii}$ and $\delta \mathcal{C}_{\text{cVHS},ii}$ never perfectly cancel, i.e. $a_{xx (yy)}+\delta a_{xx (yy)} >0$ regardless of the dispersion details. Therefore, the net peak intensity in the presence of multiple cVHS is always log divergent $\mathcal{C}^{(0)}_{\text{cVHS},ii}+\delta \mathcal{C}_{\text{cVHS},ii}\propto\log(\frac{\Lambda_i}{\sqrt{|\mu|}})$. 

The coefficient of this log divergence further depends on the isotropy of specific VHS dispersions in Eq. \ref{eq:VHS_type}. In the isotropic limit where $\alpha_1 = \beta_1$, the vertex corrections in Eq. \ref{eq:corr_multiple} vanish $\delta \mathcal{C}_{\text{cVHS},ii}=0$.  
Thus, the net peak intensity is simply given by the bare intensity $\mathcal{C}^{(0)}_{\text{cVHS},ii}$ so that the full magnitude of log divergence is kept. For anisotropic dispersions with $\alpha_1 \neq \beta_1$, the non-zero vertex corrections $\delta \mathcal{C}_{\text{cVHS},ii}$ partially cancels the bare intensity in Eq. \ref{eq:bare_cVHS} such that the coefficient of the logarithmic divergence decreases.

\begin{table}[t]
\centering
\caption{
Optical absorption signatures in $s$-wave superconducting states driven by different numbers and types of VHS on the Fermi surface. In each case, the normal state DOS exhibits a characteristic behaviour of divergence. The absorption intensity and the anisotropy ratio $\mathcal{R}\equiv \text{Re}~ \sigma_{yy}/\text{Re}~\sigma_{xx}$ are summarized.}
\label{table:iCDW}
\begin{tabular}{c|ccccc}
\noalign{\smallskip}\noalign{\smallskip}\hline\hline
~ & \; DOS \; & 
 Re $\sigma_{ii}(\omega)$ &
Anisotropy \\
\hline
Multiple cVHS  & \; Log-divergent \;  & Log-divergent & $\mathcal{R}=1$
\\
Single cVHS & \; Log-divergent \;  & Non-divergent & $\mathcal{R}\approx 1$ 
\\
Single hVHS &\; Power-law divergent \;  & Non-divergent & $\mathcal{R}\gg 1$
\\
\hline
\hline
\end{tabular}
\label{table:result}
\end{table}

In Fig. \ref{fig:plot} a, we numerically demonstrate the optical absorption spectrum calculated from Eq. \ref{eq:bare+vertex} on a square lattice 
with a dispersion  $\xi_{\tbf{k}}=-t\cos(k_x)-t\cos(k_y)-\mu$. 
The Van Hove filling is at a chemical potential of $\mu = 0$, where there are two cVHS on the Fermi surface at $\textbf{k}_{c1}=(0,\pi)$ and $\textbf{k}_{c2}=(\pi,0)$. 
Since the dispersions expanded around the two VH points are isotropic, i.e. $\alpha_1=\beta_1=t$, we numerically find that the net peak intensity fully exhibits the log-divergent peak in the bare intensity, as expected analytically from Eqs. \ref{eq:bare_cVHS} and \ref{eq:corr_multiple}.   

\begin{itemize}
    \item Single cVHS $n_{\text{VH}}=1$ 
\end{itemize}

In sharp contrast to the $n_{\text{VH}}>1$ case, we find that the vertex corrections $\delta \mathcal{C}_{\text{cVHS},ii}$ in the single cVHS case perfectly cancel the log divergent term in the bare intensity $\mathcal{C}^{(0)}_{\text{cVHS},ii}$ in Eq. \ref{eq:bare_cVHS}, i.e. $a_{xx (yy)}^{(0)}+\delta a_{xx (yy)}=0$. 
Therefore, regardless of specific VHS dispersions, the net peak intensities are not divergent. Instead, we find that the magnitudes of these non-diverging peak intensities quantitatively depend on dispersion details as 
\bea
&&
\mathcal{C}_{\text{cVHS},xx}^{(0)}+\delta \mathcal{C}_{\text{cVHS},xx} 
\approx
\frac{1}{\Lambda_x \Lambda_y}\frac{\alpha_2^2}{\sqrt{\alpha_1\beta_1}}\frac{\Lambda_x^4}{4}, 
\nn\\
&&
\mathcal{C}_{\text{cVHS},yy}^{(0)}+\delta \mathcal{C}_{\text{cVHS},yy}
\approx
\frac{1}{\Lambda_x \Lambda_y}
\frac{\beta_2^2}{\sqrt{\alpha_1\beta_1}}\frac{\Lambda_y^4}{4}. \;\;
\label{eq:total_cVHS}
\eea 
For purely quadratic cVHS dispersions with $\alpha_2=\beta_2=0$ (see Eq. \ref{eq:VHS}), the $\omega=2\Delta$ absorption peak completely vanishes since the net intensities in Eq. \ref{eq:total_cVHS} are zero.   
The vanishing peak is a consequence of current conservation due to the momentum-independent effective masses. 
For band structures where the quartic terms naturally exist $\alpha_2,\beta_2\neq0$, we find non-diverging intensities that quantitatively depend on $\alpha_2$ and $\beta_2$ (see Eq. \ref{eq:total_cVHS}).

In Fig. \ref{fig:plot}b, we numerically demonstrate this weak peak considering a rectangular lattice with a single band $\xi_{\tbf{k}}=-t_x\cos(k_x)-t_y\cos(k_y)-t_{yy}\cos(2k_y)-\mu$. The four-fold rotational symmetry is broken by the difference between the nearest-neighbor hoppings $t_x\neq t_y$ and next-nearest neighbor hopping $t_{yy}\neq 0$.  The Van Hove filling is at $\mu = 0$, where a single cVHS at $\tbf{k}_{c1}$ occurs on the Fermi surface. By expanding around $\tbf{k}_{c1}$, we can express the  cVHS dispersion coefficients in Eq. \ref{eq:VHS} in terms of the hopping integrals as $\alpha_1 = t_x$, $\beta_1 = t_y-4t_{yy}$, $\alpha_2 = -t_x/2$, and $\beta_2 = (-t_y+16t_{yy})/2$. 

Since there is only one cVHS on the Fermi surface and the dispersion near the VH point is not purely quadratic $\alpha_2$, $\beta_2\neq 0$, we analytically expect a non-diverging absorption peak at $\omega=2\Delta$ from Eq. \ref{eq:total_cVHS}.  
In the numerically obtained spectrum in In Fig. \ref{fig:plot}b, we indeed find a weak peak, where the peak intensity is two orders of magnitudes smaller than the divergent peak in the $n_{\text{VH}}=2$ cVHS case (see Fig. \ref{fig:plot}a). 




\textit {Superconductivity driven by hVHS ---} 
We consider a hVHS that is quadratic in $q_x$ but quartic along $q_y$, where the VHS dispersion in Eq. \ref{eq:VHS} has $\beta_1=0$. 
By calculating the bare and net peak intensities along the $i=x$ direction from Eq. \ref{eq:intensity}, we find 
\bea
&&
\mathcal{C}_{\text{hVHS},xx}^{(0)}
\approx 
\frac{1}{\Lambda_x \Lambda_y}
\frac{\alpha_1^{\frac{3}{2}}\Gamma(\frac{1}{4})^2}{4\sqrt{2\pi}(\frac{\beta_2}{3})^{\frac{1}{4}}}
\begin{cases}
\frac{\sqrt{2}}{\mu^{\frac{1}{4}}}
& \text{for}\;\; \mu >0 
\\
\frac{1}{|\mu^{\frac{1}{4}}|} 
& \text{for}\;\; \mu <0 
\end{cases},
\nn\\
&&
\mathcal{C}_{\text{hVHS},xx}^{(0)}
+\delta \mathcal{C}_{\text{hVHS},xx}
\approx
\frac{1}{\Lambda_x \Lambda_y}
\frac{\alpha_2^2}{7\alpha_1^{\frac{3}{4}}(\frac{\beta_2}{6})^{\frac{1}{4}}}\Lambda_x^{\frac{7}{2}}, 
\label{eq:bare_hVHS}
\eea
where $\Gamma(x)$ is the gamma function. 
Here, the power-law divergence at $\mu\rightarrow 0$ in the bare intensity $\mathcal{C}_{\text{hVHS},xx}^{(0)}$ comes from the divergent DOS and the nearly momentum $\tbf{q}$-independent effective mass $\mathbb{M}_{\tbf{q},xx}$ (see Eq. \ref{eq:intensity}). 
Nonetheless, similar to the single cVHS case in Eq. \ref{eq:total_cVHS}, this divergence is perfectly canceled by the vertex correction $\delta \mathcal{C}_{\text{hVHS},xx}$ so that the net intensity is not divergent.   
The net intensity in the $i=y$ direction is not divergent either. Specifically, we find from Eq. \ref{eq:intensity} that  
\bea
\mathcal{C}_{\text{hVHS},yy}^{(0)}+\delta \mathcal{C}_{\text{hVHS},yy}
\approx 
\frac{1}{\Lambda_x \Lambda_y}
\frac{\beta_2^2}{3(\frac{\alpha_1\beta_2}{6})^{\frac{1}{2}}}\Lambda_y^{3}
+O(|\frac{\mu}{\beta_2}|^{\frac{1}{4}}).  
\;\;\;\;\;\;
\label{eq:total_hVHS}
\eea
Different from the $i=x$ case, the lack of divergence here is due to the momentum-dependent inverse effective mass $\mathbb{M}^{-1}_{\tbf{q},yy}\sim q_y^2$ along the direction with quartic dispersion (see Eq. \ref{eq:intensity} and SM \cite{SM_optical} Sec. \ref{sec:SM_Peak_intensity}).

In Fig. \ref{fig:plot}c, we numerically demonstrate the analytically predicted non-diverging absorption peak in Eqs. \ref{eq:bare_hVHS} and \ref{eq:total_hVHS} using a lattice model with a single normal band $\xi_{\tbf{k}}=-t_x\cos(k_x)-t_y\cos(k_y)-t_{y}\cos(2k_y)/4-\mu$. This dispersion exhibits a hVHS with $\beta_1=0$ at $\textbf{k}_{c2}$, where the leading order is quartic in $k_y$ but quadratic in $k_x$.  
Consistent with our analytic results, we find that the peak intensities in both $i=x$ and $y$ directions, although not vanishing, are orders of mangitude smaller than the $n=2$ cVHS case with a diverging peak. 

\textit{Dinstinguish single cVHS and hVHS cases---}
We summarize our results for different VHS fermiology in Table \ref{table:result}. 
In short, we expect a diverging absorption peak in Re $\sigma_{ii}(\omega)$ for the multiple cVHS case, but a non-divergent or vanishing peak in cases with a single cVHS or hVHS. 
For single VHS cases with non-vanishing but order-of-magnitudes weaker peaks,   
we propose to distinguish whether the normal state contains a cVHS or hVHS qualitatively from the anisotropy between  $i=x$ and $y$ (see Fig. \ref{fig:schematic}a). 
Specifically, the single cVHS case exhibits isotropic intensities $\mathcal{C}_{\text{cVHS},xx}\sim\mathcal{C}_{\text{cVHS},yy}$ regardless of the anisotropy in hopping parameters. However, in the hVHS case, the peak anisotropy sensitively reflects the anisotropy in the hopping parameters. This sharp contrast is clearly shown in Figs. \ref{fig:plot}b and c. 

We quantify this qualitative difference by defining 
\bea
\mathcal{R}
\equiv
\text{Re}~\sigma_{yy}(\omega)/\text{Re}~\sigma_{xx}(\omega), 
\label{eq:ratio_cVHS}
\eea
where we show $\mathcal{R}$ for the single cVHS and hVHS cases in Fig. \ref{fig:plot}d as a function of hopping anisotropy $t_y/t_x$. 
For the single cVHS case, we find  $\mathcal{R}\sim 1$ regardless of the hopping anisotropy the presence of  the anisotropy in dispersion parameters (see Eq. \ref{eq:VHS}). This is because $\Lambda_y^2/\Lambda_x^2 \approx \alpha_1/\beta_1$ and $\mathcal{R}\sim\frac{(\beta_2/\beta_1)^2}{(\alpha_2/\alpha_1)^2}$ so that the dispersion anisotropy is smeared out by the ratios between quadratic and quartic coefficients $\alpha_1$ and $\alpha_2$ ($\beta_1$ and $\beta_2$) in the same direction. 
In contrast, the ratio $\mathcal{R}$ in the hVHS case is sensitive to the dispersion anisotropy. The explicit expression of $\mathcal{R}$ in terms of the dispersion coefficients is shown in SM \cite{SM_optical} Sec. \ref{sec:SM_Peak_intensity}. This is intuitively because the quadratic coefficient $\beta_1=0$ vanishes so that the dispersion anisotropy is not smeared out as in the cVHS case.

\textit {Discussion ---} Possible candidate superconductors for experimental verification of our theory include  Sr$_2$RuO$_4$ films under biaxial strain \cite{SRO_biaxial_exp,SRO_biaxial} and uniaxial strain, as well as kagome superconductors $\text{AV}_3\text{Sb}_5$ (A=K, Rb, Cs) and few-layer graphene-base superconducting systems. 
Moreover, the presence of disorders can result in a featureless background of spectral weight at a higher frequency range $\omega\leq2\Delta$ due to indirect optical transitions (see Fig. \ref{fig:schematic}a,b and SM \cite{SM_optical} Sec. \ref{sec:SM_disorder}). Therefore, we expect that the non-diverging weak peaks in the single cVHS and hVHS cares are likely to be buried in the disorder-induced background. In contrast, the diverging $2\Delta$ peak in the multiple cVHS cases can survive the disorders.

\textit {Acknowledgements ---} Y.-T.H. acknowledges support from NSF Grant No. DMR-2238748. Y.-T.H. acknowledges support from Department of Energy Basic Energy Science Award DE-SC0024291. H.-J.Y. is supported by the Society of Science Fellows Postdoctoral Program in the College of Science.


\bibliography{ref}

\setcounter{equation}{0}
\setcounter{figure}{0}
\setcounter{table}{0}
\renewcommand{\theequation}{S\arabic{equation}}
\renewcommand{\thefigure}{S\arabic{figure}}

\pagebreak
\newpage

\thispagestyle{empty}
\mbox{}
\pagebreak
\newpage
\onecolumngrid
\begin{center}
  \textbf{\large Optical absorption signatures of superconductors driven by Van Hove singularities
  \\Supplementary Materials}\\[.2cm]
  
  Hyeok-Jun Yang$^{1}$, and Yi-Ting Hsu$^1$\\[.1cm]
  {\itshape ${}^1$Department of Physics, University of Notre Dame, Notre Dame, Indiana 46556, USA\\
}
(Dated: \today)\\[1cm]
\end{center}
\onecolumngrid
These Supplementary Materials contain the details on I. Galilean invariance, II. Optical conductivity calculations, III. Peak intensity and IV. Effect of disorder.
\section{Galilean invariance}
\label{sec:SM_GI}
In this section, we summarize how the Galilean invariance is related to the absence of optical absorption in superconductors.
\subsection{Galilean transformation}
Here, we show that the system is Galilean-invariant if the effective mass is a constant tensor. This is the case of $\alpha_2=\beta_2=0$ which results in $\text{Re}\sigma_{ij}(\omega)=0$ in the main text. 

We assume that the kinetic $\hat{T}$ and interaction $\hat{V}$ terms in $\hat{H}[\hat{\tbf{x}}_\alpha,\hat{\tbf{p}}_\alpha]=\hat{T}+\hat{V}$ depend on the momentum and position operators only, i.e. $\hat{T}\equiv \hat{T}[\hat{\tbf{p}}_\alpha]$ and $\hat{V}\equiv \hat{V}[\hat{\tbf{x}}_\alpha]$, respectively. 
The Galilean transformation relates two reference frames separated by a constant velocity $\tbf{v}$,
\bea
\hat{U}_\mathcal{G}(t)^{-1}\hat{\tbf{x}}_{\alpha}\hat{U}_\mathcal{G}(t) 
= \hat{\tbf{x}}_{\alpha}+\tbf{v}t, 
\qquad
\hat{U}_\mathcal{G}(t)^{-1}\hat{\tbf{p}}_{\alpha}\hat{U}_\mathcal{G}(t) = \hat{\tbf{p}}_{\alpha}+\tbf{m}_{\alpha}\cdot\tbf{v},
\label{Seq:GT}
\eea
which can be realized by the unitary transformation (within the first quantization) \cite{gottfried2018quantum},
\bea
\hat{U}_\mathcal{G}(t) = e^{-i\tbf{v}\cdot \sum_\alpha (\hat{\tbf{p}}_{\alpha}t -\tbf{m}_\alpha \cdot \hat{\tbf{x}}_\alpha )},
\label{Seq:U_G}
\eea
where $\hat{\tbf{x}}_\alpha$ and $\hat{\tbf{p}}_\alpha$ are position and momentum operators of particle $\alpha$ which are canonically conjugate each other, $[\hat{x}_{\alpha,i},\hat{p}_{\beta,j}]=i\delta_{\alpha\beta}\delta_{ij}$. Also, the inverse of mass tensor $(\tbf{m}_\alpha)^{-1}$ is defined by the quadratic coefficients in the kinetic term,
\bea
\hat{T}=\frac{1}{2}\sum_{\alpha}\sum_{ij} (m_\alpha)^{-1}|_{ij}\hat{p}_{\alpha,i}\hat{p}_{\alpha,j}+...,
\label{Seq:T}
\eea 
where $...$ includes higher-order terms in $\hat{p}_{\alpha,i}$.
For an arbitrary wavefunction $|\psi(t)\ra$ satisfying the Schrödinger equation, $\Big(i\partial_t -\hat{H}\Big)|\psi(t)\ra=0$, the transformed wavefunction $|\psi'(t)\ra =\hat{U}_\mathcal{G}(t)|\psi(t)\ra$ should also satisfy $\Big(i\partial_t -\hat{H}\Big)|\psi'(t)\ra=0$ if the Hamiltonian $\hat{H}$ is invariant under Eq. \ref{Seq:U_G}. This results in
\bea
\hat{U}^{-1}_\mathcal{G}\hat{H}\hat{U}_\mathcal{G}=\hat{H}+i\hat{U}^{-1}_\mathcal{G}\partial_t \hat{U}_\mathcal{G}.
\label{Seq:HU}
\eea
Since $\hat{V}[\hat{\tbf{x}}_\alpha]$ depends on $\hat{\tbf{x}}_\alpha$ only, $\hat{V}[\hat{\tbf{x}}_\alpha]$ is invariant under Eq. \ref{Seq:GT} if $\hat{V}[\hat{\tbf{x}}_\alpha]$ is translational-invariant, i.e. $\hat{U}_\mathcal{G}^{-1}\hat{V}\hat{U}_\mathcal{G}=\hat{V}$.
Using Eq. \ref{Seq:U_G},
\bea
i\hat{U}_\mathcal{G}^{-1}\partial_t\hat{U}_\mathcal{G}
=
\hat{U}_\mathcal{G}^{-1}
(\tbf{v}\cdot\hat{\tbf{P}}-\frac{1}{2}\tbf{v}\cdot \tbf{M} \cdot \tbf{v} )\hat{U}_\mathcal{G} = 
\tbf{v}\cdot\hat{\tbf{P}}+\frac{1}{2} \tbf{v}\cdot \tbf{M} \cdot \tbf{v},
\eea\
where $\tbf{M}=\sum_{\alpha}\tbf{m}_\alpha$ and $\hat{\tbf{P}}=\sum_\alpha \hat{\tbf{p}}_\alpha$.
Also, 
\bea
\hat{U}_\mathcal{G}^{-1}\hat{T}\hat{U}_\mathcal{G}&=&\frac{1}{2}\sum_{\alpha} (\hat{\tbf{p}}_\alpha + \tbf{m}_\alpha \cdot \tbf{v})\cdot \tbf{m}^{-1}_{\alpha}\cdot (\hat{\tbf{p}}_\alpha + \tbf{m}_\alpha \cdot \tbf{v}) + ... \nn\\
&=&
\frac{1}{2}\sum_\alpha (\hat{\tbf{p}}_\alpha \cdot \tbf{m}^{-1}_{\alpha}\cdot \hat{\tbf{p}}_\alpha) + \tbf{v}\cdot\hat{\tbf{P}}+\frac{1}{2} \tbf{v}\cdot \tbf{M} \cdot \tbf{v} +...,
\eea
where $...$ includes higher-order terms likewise Eq. \ref{Seq:T}.
Thus the invariance, Eq. \ref{Seq:HU} holds in the absence of higher terms in Eq. \ref{Seq:T}, i.e. when $\hat{T}[\hat{\tbf{p}}_\alpha]=\frac{1}{2}\sum_\alpha \hat{\tbf{p}}_\alpha \cdot \tbf{m}^{-1}_{\alpha}\cdot \hat{\tbf{p}}_\alpha$.

\subsection{Current conservation}
The selection rule (2) in the main text indicates that the $\omega=2|\Delta|$ peak in optical absorption can only exist when the current is not conserved. This condition can be understood as follows. 
In a translational invariant system, the current change rate can be written as
\bea
\frac{d\tbf{J}(\tau)}{d\tau}
=
[H^{\text{int}},\tbf{J}(\tau)]
=
\frac{U}{2V}\sum_{\tbf{k}\tbf{k}'\tbf{p}\sigma\sigma'}c^{\dagger}_{\tbf{k}+\tbf{p}\sigma}c^{\dagger}_{\tbf{k}'-\tbf{p}\sigma'}c_{\tbf{k}'\sigma'}c_{\tbf{k}\sigma}(\Delta \tbf{v}_{\tbf{k}\tbf{k}',\tbf{p}}), \;\;
\label{eq:djdt}
\eea 
where $\tbf{J}(\tau)=e^{\tau H}\tbf{J}e^{-\tau H}$, $\tbf{J}=\frac{1}{V}\sum_{\tbf{k}\sigma}\tbf{v}_{\tbf{k}}c_{\tbf{k}\sigma}^\dagger c_{\tbf{k}\sigma}$ is the total current density and the change in velocity during a collision is given by $\Delta\tbf{v}_{\tbf{k}\tbf{k}',\tbf{p}}=\tbf{v}_{\tbf{k}+\tbf{p}}+\tbf{v}_{\tbf{k}'-\tbf{p}}-\tbf{v}_{\tbf{k}'}-\tbf{v}_{\tbf{k}}$. 
Due to momentum conservation, the change in velocity $\Delta\tbf{v}_{\tbf{k}\tbf{k}',\tbf{p}}$ vanishes for all momenta $\tbf{k}, \tbf{k}', \tbf{p}$ when the velocity is proportional to the momentum $\tbf{v}_\tbf{k}\propto\tbf{k}$. 
In such a current-conserving (i.e. Galilean symmetric) system, the current-current correlator,
\bea
P_{ij}(i\omega_m)=-\frac{1}{V}
\int_0^\beta d\tau e^{i\omega_m \tau} \la J_i(\tau) J_j(0)\ra
\eea 
vanishes with the vanishing current change rate $\frac{d\tbf{J}}{d\tau}$ since 
\bea
P_{ij}(i\omega_m)
=\frac{1}{V}\frac{1}{(i\omega_m)^2}
\int_0^\beta d\tau e^{i\omega_m \tau} \la \frac{dJ_i(\tau)}{d\tau}
\frac{dJ_j(0)}{d\tau}\ra
\eea
by performing integration by part twice. Therefore, the optical absorption vanishes Re$\sigma_{ii}(\omega)=0$ in current-conserving systems. 
The most common example occurs when the normal state has a parabolic dispersion $\epsilon_{\textbf{k}}=|\tbf{k}|^2/(2m)-\mu$, where $\textbf{k}$, $m$, and $\mu$ are the momentum, electron mass, and chemical potential, respectively. 
The velocity operator in this case is given by $\tbf{v}_\tbf{k}=\tbf{k}/m\propto\textbf{k}$ \cite{gottfried2018quantum} so that the optical absorption Re$\sigma_{ii}(\omega)=0$ \cite{https://doi.org/10.1002/andp.200651807-809}.
More generally speaking, the current is nearly conserved in systems with an effective mass $\mathbb{M}_\tbf{q}$ that is nearly momentum independent.  
Note that even in crystals with non-parabolic band structures, this condition is still commonly satisfied when the superconducting gap develops not too far from the band bottom so that the normal band is still nearly parabolic.

\section{Optical conductivity calculations}
\label{sec:SM_Optical}
This section include the calculation details for the optical absorptions in $s$-wave superconductors. Starting from the BdG (Bogoliubov-de Gennes) Hamiltonian and its vertex correction, then we apply them to the VHS dispersions.

\begin{figure}[b!]
{
\includegraphics[width=0.2\textwidth]{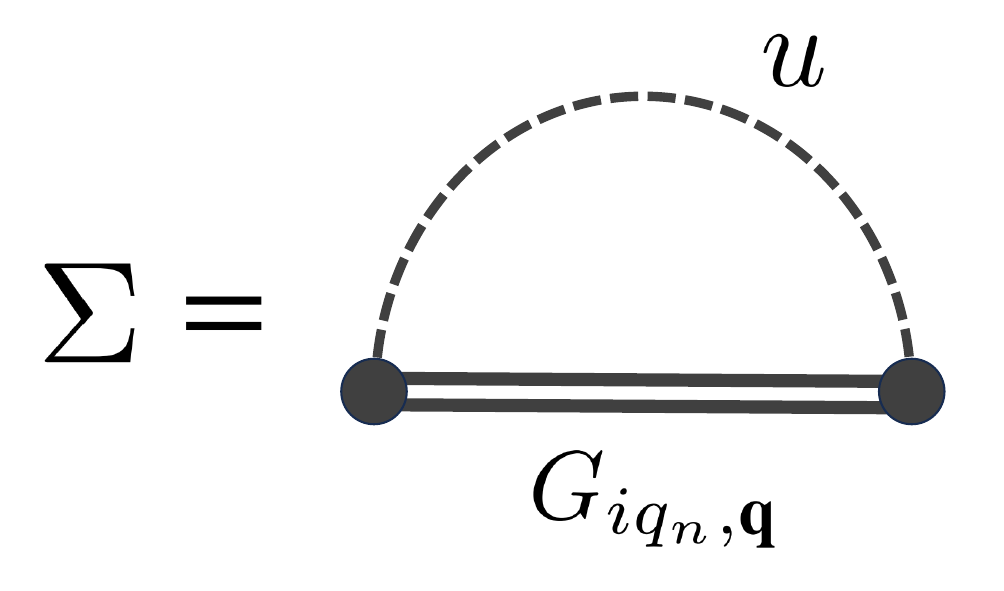}}
\caption{
Diagrammatic representation of Eq. \ref{Seq:self-energy} whose off-diagonal correction reads the gap equation. The expansion of $\Sigma$ in terms of the bare Green function includes only no-line-crossing diagrams likewise Fig. 2.
}
\label{Sfig:self-energy}
\end{figure}

\subsection{BdG Hamiltonian}
In the main text, we consider the BdG Hamiltonian when the normal state Fermi surface is entirely shifted by $\tbf{Q}/2$, 
\bea
{H}^{\text{BdG}}_{\tbf{q}}
=
\begin{pmatrix}
\xi_{\tbf{q}+\frac{\tbf{Q}}{2}} & \Delta \\
\Delta & -\xi_{-\tbf{q}+\frac{\tbf{Q}}{2}}
\end{pmatrix}
=
\delta\xi_\tbf{q}\tau^0 +
\bar{\xi}_\tbf{q}\tau^3 +\Delta \tau^1.
\label{Seq:H_BdG}
\eea
on the basis ${\Psi}_{\tbf{q}}
=
\begin{pmatrix}
c_{\tbf{q}+\tbf{Q}/2,\uparrow} \\
c_{-\tbf{q}+\tbf{Q}/2,\downarrow}^\dagger
\end{pmatrix}$ and the BdG Green function in frequency space is
\bea
G_{iq_n,\tbf{q}}=
(iq_n \tau^0 - H_{\tbf{q}}^{\text{BdG}})^{-1}=\frac{1}{(iq_n -\delta \xi_\tbf{q})^2-E_\tbf{q}^2}
\Big[
(iq_n -\delta\xi_\tbf{q})\tau^0 +
\bar{\xi}_\tbf{q}\tau^3 + \Delta \tau^{1}
\Big],
\label{Seq:BdG_Green}
\eea
where $\omega_n$ is the fermionic Matsubara frequency and $\bar{\xi}_\tbf{q} = \frac{\xi_{\tbf{q}+\tbf{Q}/2}+\xi_{-\tbf{q}-\tbf{Q}/2}}{2}$, $\delta{\xi}_\tbf{q} = \frac{\xi_{\tbf{q}+\tbf{Q}/2}-\xi_{-\tbf{q}-\tbf{Q}/2}}{2}$.
 
The $s$-wave superconducting order parameter $\Delta$ (fixed to be real) is self-consistently determined by the gap equation at $T=0$,
\bea
\Delta = -u\int_{\tbf{q}} \frac{\Delta}{2\sqrt{\bar{\xi}_{\tbf{q}}^2+|\Delta|^2}},
\label{Seq:gap_eq}
\eea
which is obtained by decoupling the full Hamiltonian, $H^{\text{full}}
=\sum_{\tbf{k}\sigma}\xi_{\tbf{k}}c^{\dagger}_{\tbf{k}\sigma}c_{\tbf{k}\sigma}
+\frac{1}{2V}\sum_{\tbf{k}\tbf{k}'\tbf{p}\sigma\sigma'}u~( c^{\dagger}_{\tbf{k}+\tbf{p}\sigma}c^{\dagger}_{\tbf{k}'-\tbf{p}\sigma'}c_{\tbf{k}'\sigma'}c_{\tbf{k}\sigma})$ with $u<0$. 
In Eq. \ref{Seq:gap_eq}, we denote $\int_{\tbf{q}}=\frac{1}{V}\sum_{\tbf{q}}$ for notational simplicity.
In the self-energy (Fig. \ref{Sfig:self-energy}), the diagonal correction is ignored, and Eq. \ref{Seq:gap_eq} is the off-diagonal correction of \cite{schrieffer2018theory, PhysRevB.95.014506}
\bea
\Sigma = T\sum_{iq_n}\int_{\tbf{q}}u~ \tau^3 G_{iq_n,\tbf{q}}\tau^3,
\label{Seq:self-energy}
\eea
which follows the same approximation scheme in the current-current correlation calculation (Fig. 2 in the main text).

For small $|\tbf{Q}|$, we have $\bar{\xi}_\tbf{q} = \xi_{\tbf{q}}+O(|\tbf{Q}|^2)$
and 
$\delta{\xi}_\tbf{q} = \frac{1}{2}\tbf{v}_{\tbf{q}}\cdot \tbf{Q} +O(|\tbf{Q}|^2)$ with $\tbf{v}_{\tbf{q}}=\nabla_{\tbf{q}}\xi_{\tbf{q}}$. Then, the eigenvalues of Eq. \ref{Seq:H_BdG} are given by,
\bea
\tilde{E}_{\tbf{q}}^{\pm}=\pm E_{\tbf{q}} + \frac{1}{2}\tbf{v}_{\tbf{q}}\cdot \tbf{Q},
\label{Seq:Bogoliubov}
\eea
where $E_{\tbf{q}}=\sqrt{\xi_{\tbf{q}}^2+|\Delta|^2}$ is the Bogoliubov dispersion in the absence of applied supercurrent $\tbf{Q}$. The magnitude of $\tbf{Q}$ is assumed to be small enough so that the spectrum Eq. \ref{Seq:Bogoliubov} is gapped for all $\tbf{q}$.

\subsection{Vertex correction}
The total current and bare current operator are $\tbf{j}=\sum_{\tbf{q}}\Psi^\dagger_{\tbf{q}} \tbf{j}_{\tbf{q}}\Psi_{\tbf{q}}$ and $\tbf{j}_\tbf{q}=-e\tbf{v}_{\tbf{q}}^{\text{BdG}}$ ($e>0$) respectively where $\tbf{v}_{\tbf{q}}^{\text{BdG}}$ is the velocity operator,
\bea
\tbf{v}_\tbf{q}^{\text{BdG}}
=\begin{pmatrix}
\tbf{v}_{\tbf{q}+\frac{\tbf{Q}}{2}} & 0 \\
0 & 
\tbf{v}_{\tbf{q}-\frac{\tbf{Q}}{2}}
\end{pmatrix}
=\bar{\tbf{v}}_\tbf{q}\tau^0 +\delta \tbf{v}_\tbf{q}\tau^3.
\label{Seq:v_BdG}
\eea
Here, $\bar{\tbf{v}}_\tbf{q}=\nabla_{\tbf{q}}\bar{\xi}_{\tbf{q}}=\nabla_\tbf{q} \xi_{\tbf{q}}+O(|\tbf{Q}|^2)
$ and 
$\delta \tbf{v}_\tbf{q}=
\nabla_{\tbf{q}} \delta\xi_{\tbf{q}}
=
\frac{1}{2}\mathbb{M}_{\tbf{q}}^{-1}\cdot \tbf{Q}+O(|\tbf{Q}|^2)$ where $(\mathbb{M}_{\tbf{q}}^{-1})_{ij}=\partial_{q_i}\partial_{q_j}\xi_{\tbf{q}}$ is the inverse effective mass. 
To calculate the current-current correlation, 
\bea
&&P_{ij}(i\omega_m)=
-T\sum_{iq_n}\int_\tbf{q}\text{Tr}
\Big[
j_{\tbf{q},i} G_{iq_n +i\omega_m,\tbf{q}}
\Gamma_{\tbf{q},j}(i\omega_m) 
G_{iq_n,\tbf{q}}\Big],\quad
\label{Seq:correlator}
\eea
the dressed current operator,
\bea
\boldsymbol{\Gamma}_{\tbf{q}}(i\omega_m)=
-e\Big(
\tbf{v}_{\tbf{q}}^{\text{BdG}}
+\delta\boldsymbol{\Gamma_{\tbf{q}}}(i\omega_m)
\Big),
\label{Seq:dressed_current}
\eea
and the vertex correction $\delta\boldsymbol{\Gamma}_{\tbf{q}}(i\omega_m)\equiv \sum_{i}(\delta\boldsymbol{\Gamma}_{\tbf{q}}(i\omega_m))_{i}\tau^i$ are introduced to satisfy the Ward identity \cite{schrieffer2018theory},
\bea
p_\mu \tilde{\Gamma}_{\tbf{q}+\tbf{p},\tbf{q}}^{\mu}(iq_n +i\omega_m,iq_n)=G^{-1}_{iq_n +i\omega_m,\tbf{q}+\tbf{p}}\tau^3 
-\tau^3 G^{-1}_{iq_n,\tbf{q}},
\label{Seq:Ward}
\eea
with $\Gamma_{\tbf{q},j}(i\omega_m)= \tilde{\Gamma}^{j}_{\tbf{q},\tbf{q}}(iq_n+i\omega_m,iq_n)$. 
Using Eqs. \ref{Seq:BdG_Green} and \ref{Seq:Ward}, we have 
$\delta\boldsymbol{\Gamma}_{\tbf{q}}(i\omega_m)=(\delta\boldsymbol{\Gamma}(i\omega_m))_2\tau^2$ for some function $(\delta\boldsymbol{\Gamma}(iq_m))_2$ \cite{PhysRevB.95.014506, PhysRevB.106.L220504} since the diagonal correction in Eq. \ref{Seq:self-energy} is ignored. Inserting this into the Bethe–Salpeter equation, 
\bea
&&\boldsymbol{\Gamma}_{\tbf{q}}(i\omega_m)
=\tbf{j}_\tbf{q}
-T\sum_{iq_n}\int_{\tbf{q}'}u\tau^3 G_{iq_n+i\omega_m,\tbf{q}'} \boldsymbol{\Gamma}_{\tbf{q}'}(i\omega_m)
G_{iq_n,\tbf{q}'}\tau^3, ~~~
\label{Seq:Bethe-Salpeter}
\eea
leads to,
\bea
&&
(\delta{\boldsymbol\Gamma}(i\omega_m))_2
=
-T\sum_{iq_n}\int_{\tbf{q}}
u
\frac{1}{(iq_n +i\omega_m -\delta \xi_{\tbf{q}})^2-E_{\tbf{q}}^2}
\frac{1}{(iq_n -\delta \xi_{\tbf{q}})^2-E_{\tbf{q}}^2}
\nn\\
&&
\qquad \qquad \qquad 
\times
\Big[
(\delta\boldsymbol{\Gamma}(i\omega_m))_2\Big(-(iq_n+i\omega_m -\delta\xi_{\tbf{q}})
(iq_n -\delta\xi_{\tbf{q}})
+E_{\tbf{q}}^2  \Big)
-i(i\omega_m \Delta \delta{\tbf{v}}_{\tbf{q}})
\Big],
\label{Seq:self-consistent_1}
\eea
where the $\tbf{q}$-dependence in $(\delta{\boldsymbol\Gamma}(i\omega_m))_2$ vanishes since the the interaction strength $u$ has no $\tbf{q}$-dependence. The Mastubara sum in Eq. \ref{Seq:self-consistent_1} can be done by
\bea
&&
T\sum_{iq_n}
\frac{1}{(iq_n+i\omega_m-\delta\xi_\tbf{q})^2-E_\tbf{q}^2}
\frac{1}{(iq_n-\delta\xi_\tbf{q})^2-E_\tbf{q}^2}
\nn\\
&& \qquad\qquad\qquad\qquad 
=
\frac{1}{4E_{\tbf{q}}^2}
\Big(n_F(\delta \xi_{\tbf{q}}+E_{\tbf{q}})-n_F(\delta \xi_{\tbf{q}}-E_{\tbf{q}}) \Big)
\Big(
\frac{1}{i\omega_m -2E_{\tbf{q}}}
-
\frac{1}{i\omega_m +2E_{\tbf{q}}}
\Big),
\nn\\
&&
T\sum_{iq_n}
\frac{iq_n +i\omega_m-\delta\xi_\tbf{q}}{(iq_n +i\omega_m-\delta\xi_\tbf{q})^2-E_\tbf{q}^2}
\frac{iq_n -\delta\xi_\tbf{q}}{(iq_n-\delta\xi_\tbf{q})^2-E_\tbf{q}^2}
,
\nn\\
&& \qquad\qquad\qquad\qquad 
=-\frac{1}{4}\Big(n_F(\delta \xi_{\tbf{q}}+E_{\tbf{q}})-n_F(\delta \xi_{\tbf{q}}-E_{\tbf{q}}) \Big)
\Big(
\frac{1}{i\omega_m -2E_{\tbf{q}}}
-
\frac{1}{i\omega_m +2E_{\tbf{q}}}
\Big),
\label{Seq:Matsubara}
\eea
where $n_F(E)=1/(e^{\beta E}+1)$ ($\beta = 1/T$) is the Fermi distribution. At $T=0$, Eq. \ref{Seq:self-consistent_1} reads,
\bea
-u\Big(
\int_{\tbf{q}}\frac{1}{2E_{\tbf{q}}}+\frac{2E_{\tbf{q}}}{(i\omega_m)^2-4E_{\tbf{q}}^2}
\Big)
(\delta{\boldsymbol\Gamma}(i\omega_m))_2
=
-u\int_{\tbf{q}}\frac{1}{E_{\tbf{q}}}\frac{1}{(i\omega_m)^2-4E_{\tbf{q}}^2}i(i\omega_m \Delta \delta\tbf{v}_{\tbf{q}}),
\label{Seq:Seq:self-consistent_2}
\eea
using Eq. \ref{Seq:gap_eq}. Then,
\bea
(\delta{\boldsymbol\Gamma}(i\omega_m))_2
= 
\frac{2i\Delta}{i\omega_m}
\frac{ I_{i\omega_m}[\delta \tbf{v}_{\tbf{q}}]}{I_{i\omega_m}[1]} 
 ,
\label{Seq:vertex_sol}
\eea 
where
\bea
I_{i\omega_m}[f_{\tbf{q}}]=\int_{\tbf{q}}\frac{1}{E_{\tbf{q}}}\frac{f_{\tbf{q}}}{(i\omega_m)^2-4E_{\tbf{q}}^2},
\eea
is a functional of $f_{\tbf{q}}$ \cite{PhysRevB.95.014506}. 

\subsection{Current-current correlation}
Here, we derive Eqs. 8 and 9 in the main text.
The bare current-current correlation and its correction are,
\bea
&&
P^{(0)}_{ij}(i\omega_m)
=
-e^2T\sum_{iq_n}\int_\tbf{q}\text{Tr}
\Big[
{v}_{\tbf{q},i}^{\text{BdG}} G_{iq_n +i\omega_m,\tbf{q}}
{v}_{\tbf{q},j}^{\text{BdG}}
G_{iq_n,\tbf{q}}\Big],
\label{Seq:P_bare}
\eea
and
\bea
&&
\delta P_{ij}(i\omega_m)
=
-e^2 T\sum_{iq_n}\int_\tbf{q}\text{Tr}
\Big[
{v}_{\tbf{q},i}^{\text{BdG}} G_{iq_n +i\omega_m,\tbf{q}}
\delta\Gamma_{\tbf{q},j}(i\omega_m) 
G_{iq_n,\tbf{q}}\Big].
\label{Seq:P_delta}
\eea
Inserting Eqs. \ref{Seq:BdG_Green} and \ref{Seq:v_BdG} into Eq. \ref{Seq:P_bare}, we have
\bea
P_{ij}^{(0)}(i\omega_m)=-4e^2 \Delta^2 I_{i\omega_m}[\delta v_{\tbf{q},i}\delta v_{\tbf{q},j}]
\label{Seq:P_bare_result}
\eea
using Eq. \ref{Seq:Matsubara}. Similarly, the correction term Eq. \ref{Seq:P_delta} becomes
\bea
\delta P_{ij}(i\omega_m)=4e^2\Delta^2 \frac{I_{i\omega_m}[\delta v_{\tbf{q},i}] I_{i\omega_m}[\delta v_{\tbf{q},j}]}{I_{i\omega_m}[1]}.
\label{Seq:P_delta_result}
\eea
The numerical calculations (Fig. 3 in the main text) are done by inserting the lattice dispersion into Eqs. \ref{Seq:P_bare_result} and \ref{Seq:P_delta_result}. If the effective mass is constant in $\tbf{q}$-space, $\delta \tbf{v}_{\tbf{q}}$ has also no $\tbf{q}$-dependence and 
$P_{ij}^{(0)}+\delta P_{ij}=0$.
  
With the analytic continuation $i\omega_m \rightarrow \omega + i\eta$, 
\bea
\text{Im}I_{\omega +i\eta}[f_{\tbf{q}}]&=&
-\frac{\pi}{\omega^2}\int_{\tbf{q}}\delta(\omega - 2E_{\tbf{q}}) f_{\tbf{q}}
=
-\frac{\pi}{4\omega \sqrt{(\frac{\omega}{2})^2 -|\Delta|^2 }}
\Theta(\omega -2|\Delta|)
\int_{\tbf{q}} 
\delta\Big(\xi_{\tbf{q}}-\sqrt{(\frac{\omega}{2})^2 -|\Delta|^2 }\Big) f_{\tbf{q}}
\nn\\
&\approx & 
-\frac{\pi}{4\omega \sqrt{(\frac{\omega}{2})^2 -|\Delta|^2 }}
\Theta(\omega -2|\Delta|)
\int_{\tbf{q}} 
\delta(\xi_{\tbf{q}}) f_{\tbf{q}},
\label{Seq:Im_I}
\eea
where the identity $\delta(F(x)-F(x_0))=\delta(x-x_0)|dF/dx|^{-1}_{x=x_0}$ is used and the last approximation holds close to the gap edge $\omega \approx 2|\Delta|$.
For $\omega \approx 2|\Delta|$, the imaginary part of $I_{\omega+i\eta}$ is much larger than the real part so Eqs. \ref{Seq:P_bare_result} and \ref{Seq:P_delta_result} can be simplified using Eq. \ref{Seq:Im_I}. Then, the optical absorption $\text{Re}~\sigma_{ij}=\sigma_{ij}^{(0)}+\delta \sigma_{ij}$ is 
\bea
&&
\sigma_{ij}^{(0)}(\omega)=\text{Im}\frac{P_{ij}^{(0)}(\omega +i\eta)}{\omega}
=\Theta(\omega -2|\Delta|)\frac{e^2\pi |\Delta|^2}{2\omega^2 \sqrt{\omega^2 -4|\Delta|^2}}
\sum_{i'j'}\tilde{\mathcal{C}}^{(0)}_{ii',jj'}
Q_{i'}Q_{j'},
\label{Seq:sigma_bare}
\\
&&
\delta\sigma_{ij}(\omega)=\text{Im}\frac{\delta P_{ij}(\omega +i\eta)}{\omega}
=
\Theta(\omega -2|\Delta|)\frac{e^2\pi |\Delta|^2}{2\omega^2 \sqrt{\omega^2 -4|\Delta|^2}}
\sum_{i'j'}\delta\tilde{\mathcal{C}}_{ii',jj'}Q_{i'}Q_{j'},\quad
\label{Seq:sigma_delta}
\eea
where the peak intensities of optical absorption,
\bea
&&
\tilde{\mathcal{C}}^{(0)}_{ii',jj'}=\int_{\tbf{q}}\delta(\xi_{\tbf{q}})
(\mathbb{M}^{-1}_{\tbf{q},ii'})
(\mathbb{M}^{-1}_{\tbf{q},jj'}),
\qquad
\delta \tilde{\mathcal{C}}_{ii',jj'}
=
-
\frac{\int_{\tbf{q}}\delta(\xi_{\tbf{q}})
(\mathbb{M}^{-1}_{\tbf{q},ii'})
\int_{\tbf{q}}\delta(\xi_{\tbf{q}})
(\mathbb{M}^{-1}_{\tbf{q},jj'})}{\int_{\tbf{q}}\delta(\xi_{\tbf{q}})},
\label{Seq:intensity_def}
\eea 
are determined by the integral of inverse effective mass along the normal state Fermi surface. In the main text, we consider the case of $\mathbb{M}^{-1}_{\tbf{q},xy}=0$ with $\tbf{Q}=Q \hat{x}$ or $\tbf{Q}=Q\hat{y}$, then $\mathcal{C}_{ii}^{(0)}=\tilde{\mathcal{C}}_{ii,ii}^{(0)}, \delta{\mathcal{C}}_{ii}=\delta\tilde{\mathcal{C}}_{ii,ii}$ in Eq. 9.

\section{Peak intensity}
\label{sec:SM_Peak_intensity}
Here, Eqs. 10-12 are derived and the expressions of non-diverging parts are shown.
\subsection{Multiple cVHS dispersion}
When the system preserves the $n$-fold rotational symmetry $\mathcal{R}_n$, there are multiple VHS at $\tbf{M}, \mathcal{R}_n \tbf{M}, \mathcal{R}_n^2 \tbf{M},...$ at the Fermi surface. 
In this case, the sum of effective mass at $\tbf{M}, \mathcal{R}_n \tbf{M}, \mathcal{R}_n^2 \tbf{M},...$ cancels each other and $\int_{\tbf{q}}\delta(\xi_{\tbf{q}})
(\mathbb{M}^{-1}_{\tbf{q},ij})$ in the vertex correction Eq. \ref{Seq:sigma_delta} is small compared to the bare intensity. For example, for the square lattice dispersion,
\bea
\xi_{\tbf{k}}=-t\cos(k_x)-t\cos(k_y)-t'\cos(2k_x)-t'\cos(2k_y)-\mu,
\label{Seq:cVHS_square}
\eea
there are two-cVHS 
at $\tbf{M}_1=(0,\pi)$ and $\tbf{M}_2=(\pi,0)$ as long as $|t|>4|t'|$. 
Close to $\tbf{M}_{1,2}$, the continuum dispersions are 
\bea
\xi_{\tbf{q}=\tbf{k}-\tbf{M}_1}
=
\frac{\alpha}{2}q_x^2-\frac{\beta}{2}q_y^2+O(q_x^4,q_y^4),
\qquad
\xi_{\tbf{q}=\tbf{k}-\tbf{M}_2}=-\frac{\beta}{2}q_x^2+\frac{\alpha}{2}q_y^2+O(q_x^4,q_y^4)
\label{Seq:cVHS_square_continuum}
\eea
where $\alpha = t+4t', \beta = t-4t', \mu =-2t'$ and these dispersions are related by $\mathcal{R}_4$ rotational symmetry.

When $t'=0$, then $\alpha=\beta$ which results in $\int_\tbf{q} \delta(\xi_{\tbf{q}})(\mathbb{M}^{-1}_{\tbf{q},ij})=0$ and $\delta \sigma_{ij}=0$. 
Due to the $\mathcal{R}_4$ rotational symmetry, $\sigma_{xx}=\sigma_{yy}$ and its intensity can be calculated using Eq. \ref{Seq:cVHS_square},
\bea
\mathcal{C}^{(0)}_{xx}
=
\int_{\tbf{k}} \delta(\xi_{\tbf{k}})(\mathbb{M}^{-1}_{\tbf{k},xx})^2
=
4t\int_{\Lambda_0}^{\pi-\Lambda_0}\frac{dk_xdk_y}{(2\pi)^2} \delta(k_x-k_y+\pi) \frac{\cos^2 k_x}{|\sin k_x|}
\approx\frac{2t}{\pi^2}\log \frac{1}{\Lambda_0},
\label{Seq:cVHS_multiple_intensity}
\eea
where the diverging integral is regulated by the long wavelength cut-off $\Lambda_0 =O(1/L)$ with the linear system size $L$.
In this case, the log-divergence in Eq. \ref{Seq:cVHS_multiple_intensity} and Eq. \ref{Seq:sigma_bare} are not cancelled by the vertex correction and the peak intensity of $\text{Re}\sigma_{ii}$ is very large (Fig. 3a in the main text). 

The log-divergence of bare intensity still survives for $t' \neq 0$. 
Using Eqs. \ref{Seq:intensity_def} and \ref{Seq:cVHS_square_continuum}, the bare intensity and vertex correction are
\bea
\mathcal{C}_{xx}^{(0)} 
&\approx &
\frac{1}{\pi^2} \frac{\alpha^2+\beta^2}{\sqrt{\alpha\beta}}\log \frac{1}{\Lambda_0},
\qquad
\delta \mathcal{C}_{xx} \approx -\frac{1}{\pi^2}\frac{(\alpha -\beta)^2}{2\sqrt{\alpha \beta}}\log \frac{1}{\Lambda_0}.
\label{Seq:cVHS_square_int}
\eea
For $\alpha,\beta>0$, the log-divergence is not perfectly cancelled in the net intensity,
\bea
\mathcal{C}_{xx}^{(0)} 
+\delta \mathcal{C}_{xx} \approx
\frac{1}{\pi^2}
\frac{(\alpha +\beta)^2}{2\sqrt{\alpha\beta}}
\log \frac{1}{\Lambda_0}.
\label{Seq:cVHS_sqaure_net}
\eea
Finally, we check the net intensity when the system preserves the 3-fold rotational symmetry, e.g. kagome or honeycomb lattice. We consider $\mathcal{R}_3$-related three VH points at $\tbf{M}_1=(0,\frac{2\pi}{\sqrt{3}}), \tbf{M}_2=(\pi, \frac{\pi}{\sqrt{3}}), \tbf{M}_3=(-\pi,\frac{\pi}{\sqrt{3}})$ \cite{Nandkishore2012} whose dispersions upto the quadratic orders are
\bea
&&
\xi_{\tbf{q}=\tbf{k}-\tbf{M}_1}=\frac{\alpha}{2}q_x^2 -\frac{\beta}{2}q_y^2,
\qquad \quad
\xi_{\tbf{q}=\tbf{k}-\tbf{M}_2}
=
\frac{1}{2}(\frac{\alpha}{4}-\frac{3\beta}{4})q_x^2
-(\frac{\sqrt{3}\alpha}{2}+\frac{\sqrt{3}\beta}{2})q_xq_y 
+\frac{1}{2}(\frac{3\alpha}{4}-\frac{\beta}{4})q_y^2,
\nn\\
&&
\xi_{\tbf{q}=\tbf{k}-\tbf{M}_3}
=
\frac{1}{2}(\frac{\alpha}{4}-\frac{3\beta}{4})q_x^2
+(\frac{\sqrt{3}\alpha}{2}+\frac{\sqrt{3}\beta}{2})q_xq_y 
+\frac{1}{2}(\frac{3\alpha}{4}-\frac{\beta}{4})q_y^2.
\label{Seq:cVHS_hexagon_continuum}
\eea
The bare intensity and vertex correction for Eq. \ref{Seq:cVHS_hexagon_continuum} are
\bea
\mathcal{C}_{xx}^{(0)}
\approx
\frac{1}{V_{\text{BZ}}/4}\frac{9\alpha^2 +9\beta^2 -6\alpha\beta}{8\sqrt{\alpha\beta}}
\log \frac{1}{\Lambda_0},
\qquad
\delta \mathcal{C}_{xx} 
\approx
-\frac{1}{V_{\text{BZ}}/4}
\frac{3(\alpha-\beta)^2}{4\sqrt{\alpha\beta}}
\log \frac{1}{\Lambda_0},
\label{Seq:cVHS_hexagon_int}
\eea
and the net intensity is
\bea
\mathcal{C}_{xx}^{(0)} 
+\delta \mathcal{C}_{xx} \approx
\frac{1}{V_{\text{BZ}}/4}
\frac{3(\alpha +\beta)^2}{8\sqrt{\alpha\beta}}
\log \frac{1}{\Lambda_0},
\label{Seq:cVHS_hexagon_net}
\eea
where $V_{\text{BZ}}=8\pi^2/3$ is the area of the first Brillouin zone.
For multiple cVHS, the imperfect cancellation of log-divergence is due to the differences of effective masses at different $n_{\text{VH}}$-VH points (Eqs. \ref{Seq:cVHS_square_continuum} and \ref{Seq:cVHS_hexagon_continuum}) and holds regardless of lattice geometry (Eqs. \ref{Seq:cVHS_sqaure_net} and \ref{Seq:cVHS_hexagon_net}).
In the next subsection, it turns out that if the single VH point is isolated, the log-divergence in bare intensity is perfectly cancelled by the vertex correction.

\subsection{Single cVHS dispersion}
Close to the VH point, the continuum dispersion upto the quartic order is
\bea
\xi_{\tbf{q}}=\Big(\frac{\alpha_1}{2}q_x^2+
\frac{\alpha_2}{12}q_x^4\Big)
-\Big(
\frac{\beta_1}{2}q_y^2+
\frac{\beta_2}{12}q_y^4
\Big)+\frac{\gamma}{2}q_x^2q_y^2 -\mu +O(q_x^6,q_y^6),
\label{Seq:VHS_dispersion} 
\eea
given that $\xi_{\tbf{q}}$ is symmetric under $q_x \rightarrow -q_x$ and $q_y \rightarrow -q_y$. This dispersion is defined within a finite region, i.e. $|q_{x,y}|<\Lambda_{x,y}$ and $\int_{\tbf{q}}=\frac{1}{\Lambda_x\Lambda_y}\int_{0}^{\Lambda_x}dq_x\int_{0}^{\Lambda_y}dq_y$.
The coefficients in Eq. \ref{Seq:VHS_dispersion} are determined by the hopping strengths and lattice geometry.
The inverse effective mass of Eq. \ref{Seq:VHS_dispersion} is
\bea
\mathbb{M}^{-1}_{\tbf{q}}=
\begin{pmatrix}
\alpha_1 +\alpha_2 q_x^2+\gamma q_y^2 & 2\gamma q_xq_y \\ 2\gamma q_xq_y &
-\beta_1 -\beta_2 q_y^2+\gamma q_x^2
\end{pmatrix}.
\label{Seq:VHS_effective_mass}
\eea
For cVHS, the quadratic orders are leading terms in Eq. \ref{Seq:VHS_dispersion} with $\alpha_1, \beta_1 \gg |\alpha_2|, |\beta_2|, |\gamma|$ and the patch widths are controlled by $\alpha_1, \beta_1$,
\bea
\alpha_1 \Lambda_x^2 \approx \beta_1 \Lambda_y^2.
\label{Seq:cVHS_Lambda}
\eea
Also,
\bea
\delta(\xi_{\tbf{q}})
&\approx &
\delta
\Big(\frac{\alpha_1}{2}q_x^2-\frac{\beta_1}{2}q_y^2-\mu \Big) 
\nn\\
&=& 
\frac{\Theta(q_y^2+\frac{2\mu}{\beta_1})}{\sqrt{\alpha_1\beta_1}\sqrt{q_y^2+\frac{2\mu}{\beta_1}}}\delta\Big(q_x-\sqrt{\frac{\beta_1}{\alpha_1}(q_y^2+\frac{2\mu}{\beta_1})}\Big)
=
\frac{\Theta(q_x^2-\frac{2\mu}{\alpha_1})}{\sqrt{\alpha_1\beta_1}\sqrt{q_x^2-\frac{2\mu}{\alpha_1}}}\delta\Big(q_y-\sqrt{\frac{\alpha_1}{\beta_1}(q_x^2-\frac{2\mu}{\alpha_1})}\Big).
\label{Seq:cVHS_delta}
\eea
and
\bea
\int_{\tbf{q}}\delta(\xi_{\tbf{q}})\approx
\frac{1}{\tilde{\Lambda}^2}\log\Big(\frac{\tilde{\Lambda}}{\sqrt{|\mu|/2}} \Big).
\label{Seq:cVHS_1}
\eea
where $\tilde{\Lambda}=\sqrt{\alpha_1}\Lambda_x$ and small $|\mu| \ll \tilde{\Lambda}^2$.
Also, we can calculate Eqs. 10 and 11 in the main text,
\bea
\mathcal{C}_{xx}^{(0)}
=
\int_{\tbf{q}}\delta(\xi_{\tbf{q}}) (\mathbb{M}_{\tbf{q},xx}^{-1})^2
&\approx &
\frac{1}{\tilde{\Lambda}^2}
\Big[
(\alpha_1 -\frac{2\gamma \mu}{\beta_1}
)^2
\log\Big(\frac{\tilde{\Lambda}}{\sqrt{|\mu|/2}}\Big)
+
2(\alpha_1 -\frac{2\gamma \mu}{\beta_1}
)(\alpha_2 +\frac{\gamma\alpha_1}{\beta_1})\frac{1}{2}\Lambda_x^2 
+
(\alpha_2 +\frac{\gamma\alpha_1}{\beta_1})^2 \frac{1}{4}\Lambda_x^4
\Big],\qquad
\label{Seq:cVHS_int_mxx2}
\\
\mathcal{C}_{yy}^{(0)}
=
\int_{\tbf{q}}\delta(\xi_{\tbf{q}})(\mathbb{M}_{\tbf{q},yy}^{-1})^2
& \approx &
\frac{1}{\tilde{\Lambda}^2}
\Big[
(\beta_1-\frac{2\gamma\mu}{\alpha_1})^2
\log\Big(\frac{\tilde{\Lambda}}{\sqrt{|\mu|/2}}\Big)
+2(\beta_1-\frac{2\gamma\mu}{\alpha_1})
(\beta_2-\frac{\gamma\beta_1}{\alpha_1})\frac{1}{2}\Lambda_y^2
+(\beta_2-\frac{\gamma\beta_1}{\alpha_1})^2\frac{1}{4}\Lambda_y^4
\Big],
\label{Seq:cVHS_int_myy2}
\eea
and
\bea
\int_{\tbf{q}}\delta(\xi_{\tbf{q}}) (\mathbb{M}_{\tbf{q},xx}^{-1})
&\approx &
\frac{1}{\tilde{\Lambda}^2}
\Big[
(\alpha_1 
 -\frac{\gamma \mu}{\beta_1}
)
\log\Big(\frac{\tilde{\Lambda}}{\sqrt{|\mu|/2}}\Big)
+
(\alpha_2 +\frac{\gamma\alpha_1}{\beta_1})
\frac{1}{2}\Lambda_x^2
\Big],
\label{Seq:cVHS_int_mxx1}
\\
\int_{\tbf{q}}\delta(\xi_{\tbf{q}})(\mathbb{M}_{\tbf{q},yy}^{-1})
& \approx &
\frac{1}{\tilde{\Lambda}^2}
\Big[
(\beta_1-\frac{2\gamma\mu}{\alpha_1})
\log\Big(\frac{\tilde{\Lambda}}{\sqrt{|\mu|/2}}\Big)
+(\beta_2-\frac{\gamma\beta_1}{\alpha_1})\frac{1}{2}\Lambda_y^2
\Big],
\label{Seq:cVHS_int_myy1}
\eea
thus, the net intensities are 
\bea
\mathcal{C}_{xx}^{(0)}+\delta \mathcal{C}_{xx} 
& \approx & \frac{1}{\tilde{\Lambda}^2}
(\alpha_2 +\frac{\gamma\alpha_1}{\beta_1})^2 \frac{1}{4}\Lambda_x^4
,\qquad
\mathcal{C}_{yy}^{(0)}+\delta \mathcal{C}_{yy} 
 \approx  \frac{1}{\tilde{\Lambda}^2}(\beta_2-\frac{\gamma\beta_1}{\alpha_1})^2\frac{1}{4}\Lambda_y^4.
\label{Seq:cVHS_net_int}
\eea
which reduces to Eq. 12 when $\gamma =0$.
Since the effective mass is almost constant close to VH point, the log-divergence of bare intensities (Eqs. \ref{Seq:cVHS_int_mxx2} and \ref{Seq:cVHS_int_myy2}) are perfectly cancelled by vertex corrections (Eqs. \ref{Seq:cVHS_int_mxx1} and \ref{Seq:cVHS_int_myy1}).
In the main text, we consider the case $\gamma =0$ and 
\bea
\frac{\text{Re}\sigma_{yy}(\omega)}{\text{Re}\sigma_{xx}(\omega)}
=
\frac{\mathcal{C}_{yy}^{(0)}+\delta \mathcal{C}_{yy}}{\mathcal{C}_{xx}^{(0)}+\delta \mathcal{C}_{xx}} \approx \frac{\beta_2^2\Lambda_y^4}{\alpha_2^2\Lambda_x^4} \approx 
\Big(
\frac{\beta_2/\beta_1}{\alpha_2/\alpha_1}\Big)^2,
\eea
using Eq. \ref{Seq:cVHS_Lambda}.

\subsection{Single hVHS dispersion}
Here, Eqs. (13)-(14) are derived. 
To evaluate Eq. (\ref{Seq:intensity_def}), we use the following table,
\bea
&&
\int_0^{\infty} du 
\frac{u^b}{(u^n+1)^a}
=\frac{1}{n}B(a-\frac{b+1}{n},\frac{b+1}{n})
,\qquad
\int_1^{\infty} du \frac{u^b}{(u^n-1)^a}=\frac{1}{n}B(1-a,a-\frac{b+1}{n}),
\label{Seq:Beta}
\eea
for $na>b+1$ where
\bea
B(p,q)=\int_0^1 t^{p-1}(1-t)^{q-1}dt,
\label{Seq:Beta_def}
\eea
is the Beta function.

At hVHS, we consider $\beta_1=0 ,\beta_2 \gg |\gamma|$ and $\alpha_1 \gg |\alpha_2|,|\gamma|$ in Eq. \ref{Seq:VHS_dispersion}. Similar to Eqs. \ref{Seq:cVHS_Lambda}-\ref{Seq:cVHS_1}, we have
\bea
\alpha_1\Lambda_x^2 
\approx
 \frac{\beta_2}{6}\Lambda_y^4,
\label{Seq:hVHS_Lambda}
\eea
and the delta function is approximated as
\bea
\delta(\xi_{\tbf{q}}) 
& \approx & 
\delta\Big(\frac{\alpha_1}{2} q_x^2-\frac{\beta_2}{12} q_y^4-\mu \Big)
\nn\\
&=&
\frac{\Theta(\frac{\beta_2}{6} q_y^4+2\mu)}{\sqrt{\alpha_1}\sqrt{\frac{\beta_2}{6} q_y^4+2\mu}}\delta\Big(q_x-\sqrt{\frac{\frac{\beta_2}{6} q_y^4+2\mu}{\alpha_1}}\Big)
=\frac{\Theta(6\alpha_1 q_x^2-12\mu)}{\frac{1}{3}\beta_2^{\frac{1}{4}}(6\alpha_1 q_x^2-12\mu)^{\frac{3}{4}}}
\delta\Big(q_y-(\frac{6\alpha_1q_x^2-12\mu}{\beta_2})^{\frac{1}{4}}\Big),
\label{Seq:hVHS_delta}
\eea
and
\bea
\int_{\tbf{q}}\delta(\xi_{\tbf{q}})\approx 
D_{\text{sgn}(\mu)}\frac{1}{|\mu|^\frac{1}{4}}, 
\qquad
\Big(D_+ =
\frac{1}{\Lambda_x\Lambda_y}\frac{\Gamma(\frac{1}{4})^2}{4\sqrt{\pi}\alpha_1^{\frac{1}{2}}(\frac{\beta_2}{3})^{\frac{1}{4}}} 
,\;\;
D_-=
\frac{D_+}{\sqrt{2}}
\Big).
\label{Seq:hVHS_1}
\eea
And the integrals of inverse effective mass (Eqs. 13 and 14 in the main text) are
\bea
\mathcal{C}_{xx}^{(0)}
&=&
\int_{\tbf{q}}\delta(\xi_{\tbf{q}}) (\mathbb{M}_{\tbf{q},xx}^{-1})^2
\approx 
\frac{1}{\Lambda_x\Lambda_y}
\Big[
\alpha_1^2 D_{\text{sgn}(\mu)}\frac{1}{|\mu|^{\frac{1}{4}}}
\nn\\
&& \qquad\qquad
+
\frac{1}{\frac{1}{3}(6\alpha_1)^\frac{3}{4}\beta_2^{\frac{1}{4}}}
\Big(2\alpha_1\gamma\sqrt{\frac{6\alpha_1}{\beta_2}}2\Lambda_x^{\frac{1}{2}}
+(2\alpha_1\alpha_2 +\frac{6\gamma^2\alpha_1}{\beta_2})\frac{2\Lambda_x^{\frac{3}{2}}}{3}
+2\alpha_1\gamma\sqrt{\frac{6\alpha_1}{\beta_2}}\frac{2\Lambda_x^{\frac{5}{2}}}{5}
+\frac{2\alpha_2^2\Lambda_x^{\frac{7}{2}}}{7}
\Big)\Big], \qquad
\label{Seq:hVHS_int_mxx2}
\\
\mathcal{C}_{yy}^{(0)}
&=&
\int_{\tbf{q}}\delta(\xi_{\tbf{q}}) (\mathbb{M}_{\tbf{q},yy}^{-1})^2
\approx 
\frac{1}{\Lambda_x\Lambda_y}
\frac{1}{\sqrt{\alpha_1\beta_2/6}} 
\Big[
\beta_2^2\frac{\Lambda_y^3}{3}
-
\frac{\gamma \beta_2^2}{3\alpha_1}\frac{\Lambda_y^5}{5}
+\frac{\gamma^2\beta_2^2}{36\alpha_1^2}\frac{\Lambda_y^7}{7}
\Big],
\label{Seq:hVHS_int_myy2}
\eea
and
\bea
\int_{\tbf{q}}\delta(\xi_{\tbf{q}}) (\mathbb{M}_{\tbf{q},xx}^{-1})
& \approx &
\frac{1}{\Lambda_x\Lambda_y}
\Big[
\alpha_1 D_{\text{sgn}(\mu)}\frac{1}{|\mu|^{\frac{1}{4}}}
+\frac{1}{\frac{1}{3}(6\alpha_1)^\frac{3}{4}\beta_2^{\frac{1}{4}}}
\Big(
\gamma\sqrt{\frac{6\alpha_1}{\beta_2}}2\Lambda_x^{\frac{1}{2}}
+\alpha_2 \frac{2\Lambda_x^{\frac{3}{2}}}{3}
\Big)
\Big],
\label{Seq:hVHS_int_mxx1}
\\
\int_{\tbf{q}}\delta(\xi_{\tbf{q}}) (\mathbb{M}_{\tbf{q},yy}^{-1})
& \approx &
\frac{1}{\Lambda_x\Lambda_y}
\frac{1}{\sqrt{\alpha_1\beta_2/6}} 
\Big[
-\beta_2\Lambda_y +\frac{\gamma\beta_2}{6\alpha_1}\frac{\Lambda_y^3}{3}
\Big],
\label{Seq:hVHS_int_myy1}
\eea
thus the bare intensity $\mathcal{C}_{yy}^{(0)}$ has no divergence and the vertex correction along $y$-direction is negligible $\delta \mathcal{C}_{yy}\approx 0$. Meanwhile, 
\bea
\mathcal{C}_{xx}^{(0)}+\delta \mathcal{C}_{xx}
\approx
\frac{1}{\Lambda_x\Lambda_y}
\frac{1}{\frac{1}{3}(6\alpha_1)^\frac{3}{4}\beta_2^{\frac{1}{4}}}
\Big(\frac{4\gamma^2\alpha_1}{\beta_2}\Lambda_x^{\frac{3}{2}}
+\frac{4\alpha_1\gamma}{5}\sqrt{\frac{6\alpha_1}{\beta_2}}\Lambda_x^{\frac{5}{2}}+\frac{2\alpha_2^2}{7}\Lambda_x^{\frac{7}{2}}
\Big).
\label{Seq:hVHS_int_net}
\eea
the power-divergence in Eq. \ref{Seq:hVHS_int_mxx2} is perfectly cancelled by the vertex correction, Eq. \ref{Seq:hVHS_int_mxx1}.

\section{Effect of disorder}
\label{sec:SM_disorder}
\begin{figure}[b!]
{
\includegraphics[width=0.2\textwidth]{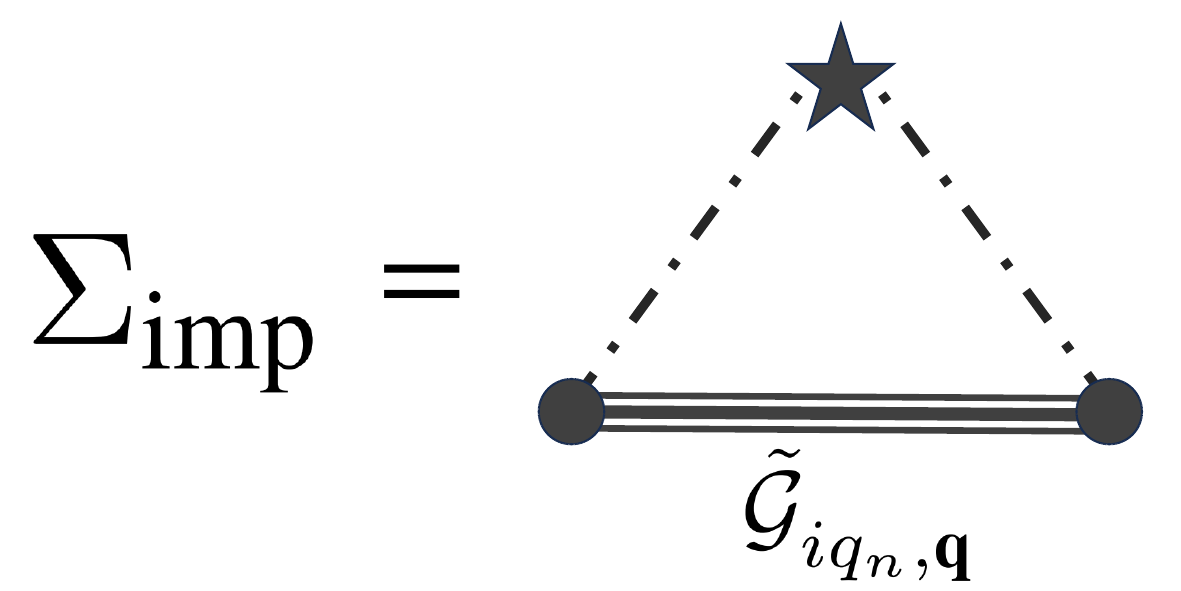}}
\caption{
Diagrammatic representation of Eq. \ref{Seq:Born_approx}. The dotted line associated with a star represents the impurity potential, Eq. \ref{Seq:H_imp}.
}
\label{Sfig:Born_approx}
\end{figure}

\begin{figure}[t!]
{\includegraphics[width=0.4\textwidth]{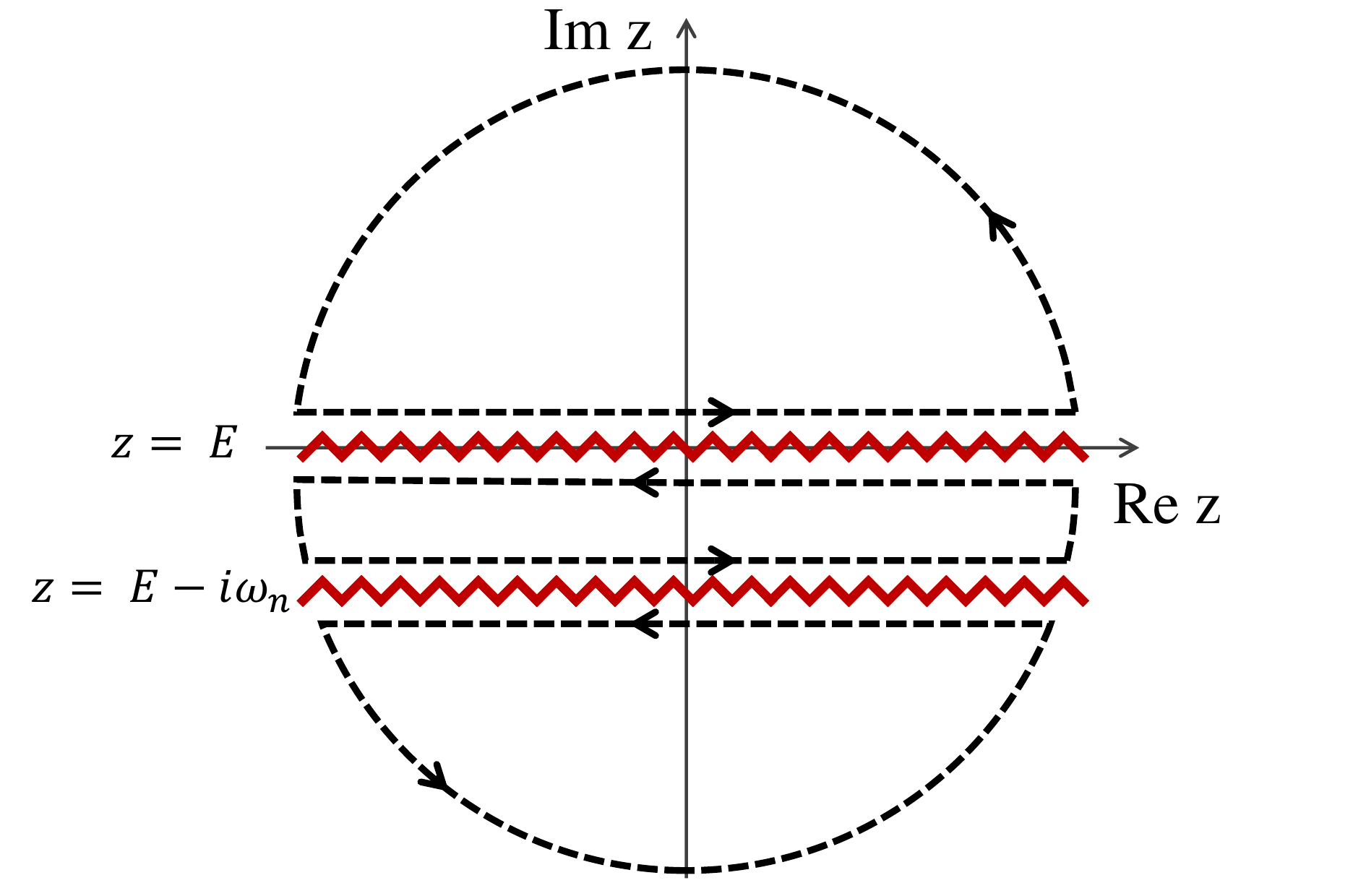}}
\caption{The contour $\mathcal{C}$ around two branch cuts $z=E$ and $z=E-i\omega_m$ to calculate the Mastubara sum in Eq. \ref{Seq:Born_Corr}.
}
\label{Sfig:Contour}
\end{figure}

In real materials, the disorder-mediated response coexist with the intrinsic response. 
In the presence of disorder, the gap-edge peak (Eqs. \ref{Seq:sigma_bare} and \ref{Seq:sigma_delta}) is smeared and can be quantitatively screened by featureless background depending on control parameters. 
The local potential term for the non-magnetic impurity is
\bea
H_{\text{imp}}=\sum_{\tbf{r}\sigma} V(\tbf{r})
c^\dagger_{\sigma}({\tbf{r}}) c_{\sigma}(\tbf{r})
\label{Seq:H_imp}
\eea
where $c^\dagger_{\sigma}(\tbf{r})$ creates an electron of spin $\sigma$ at site $\tbf{r}$ and 
\bea
\la V(\tbf{r})\ra_{\text{imp}} =0, \qquad
\la V(\tbf{r})V(\tbf{r}')\ra_{\text{imp}} = n_{\text{imp}} u_0^2 \delta(\tbf{r}-\tbf{r}'),
\label{Seq:V_imp}
\eea
where $\la ... \ra_{\text{imp}}$ is the average over impurity sites, $u_0$ is the potential strength and $n_{\text{imp}}$ is the impurity density. 
The self-energy (Fig. \ref{Sfig:Born_approx}) in addition to the BdG Green function Eq. \ref{Seq:BdG_Green} can be obtained by the self-consistent Born approximation \cite{bruus2004many}, 
\bea
\Sigma_{\text{imp}}(iq_n)
=
n_{\text{imp}} u_0^2
\int_{\tbf{q}}\tau^3 
{\tilde{\mathcal{G}}}_{iq_n,\tbf{q}}
\tau^3.
\label{Seq:Born_approx}
\eea
As a result, the frequency $z=iq_n -\delta\xi_\tbf{q}$ and the pairing gap $\Delta$ are renormalized in the full Green function $\tilde{\mathcal{G}}_{iq_n,\tbf{q}}^{-1}={G}_{iq_n,\tbf{q}}^{-1}-\Sigma_{\text{imp}}(iq_n)$,
\bea
\tilde{\mathcal{G}}_{iq_n,\tbf{q}}^{-1}=(i\tilde{q}_n-\delta\xi_{\tbf{q}})\tau^0 -{\xi}_{\tbf{q}}\tau^3 -\tilde{\Delta}\tau^1,
\label{Seq:Born_Green_function}
\eea
where
\bea
&&
\tilde{z}(z)
=z+in_\text{imp}u_0^2~\Gamma_1(\tilde{z},\tilde{\Delta})
=
z-n_\text{imp}u_0^2
\int_\tbf{q}\frac{\tilde{z}}{\tilde{z}^2-\xi^2_\tbf{q}-\tilde{\Delta}^2}
,
\nn\\
&&
\tilde{\Delta}(z)
=
\Delta +in_\text{imp}u_0^2~\Gamma_2(\tilde{z},\tilde{\Delta})
=
\Delta
-n_\text{imp}u_0^2
\int_\tbf{q}\frac{\tilde{\Delta}}{\tilde{z}^2-\xi^2_\tbf{q}-\tilde{\Delta}^2}
.
\label{Seq:renormalized}
\eea
Since the real part of $\Sigma_{\text{imp}}(iq_n \rightarrow \omega +i\eta)$ yields energy shifts, we will only focus on the imaginary part. For large frequency $z$, both corrections are small, $\Gamma_1 \approx 0$ and  $\Gamma_2 \approx 0$.
The current-current correlation is (without the vertex correction),
\bea
P_{ij}(i\omega_m)=-2e^2T\sum_{iq_n}\int_\tbf{q}
v_{\tbf{q},i}v_{\tbf{q},j}
\frac{\tilde{z}\tilde{z}_++\xi_\tbf{q}^2+\tilde{\Delta}\tilde{\Delta}_+}{(\tilde{z}^2-\xi_\tbf{q}^2-\tilde{\Delta}^2)(\tilde{z}_+^2-\xi_\tbf{q}^2-\tilde{\Delta}_+^2)}
\label{Seq:Born_Corr}
\eea
where $\tilde{z}_+=\tilde{z}(z_+)$, $\tilde{\Delta}_+=\tilde{\Delta}(z_+)$ and $z_+=z+i\omega_m$.
Using the contour integral, $\text{Im}P_{ij}(\omega+i\eta)$ can be simplified for large $\omega$. Each term in Eq. \ref{Seq:Born_Corr} can be written as
\bea
\mathcal{S}(i\omega_m)
=
-T\sum_{iq_n}f_1(iq_n)f_2(iq_n +i\omega_m)
=
\oint_{\mathcal{C}} \frac{dz}{2\pi i}n_F(z)f_1(z)f_2(z+i\omega_m),
\eea
for some functions $(f_{1,2}(E+i\eta))^*=f_{1,2}(E-i\eta)$ and the contour surrounds the branch cut at $z=E$ and $z=E-i\omega_m$ shown in Fig. \ref{Sfig:Contour},
\bea
\mathcal{S}(i\omega_m)
&=&
\int_{-\infty}^{\infty} \frac{dE}{2\pi i}
\Big[
n_F(E)
\Big(f_1(E+i\eta)f_2(E+i\omega_m)-f_1(E-i\eta)f_2(E+i\omega_m)
\nn\\
&&\qquad
\qquad
+f_1(E-i\omega_m)f_2(E+i\eta)-f_1(E-i\omega_m)f_2(E-i\eta)
\Big]
\eea
and
\bea
\text{Im}\mathcal{S}(\omega+i\eta)
=\int_{-\infty}^\infty
\frac{dE}{\pi}
\Big(
n_F(E)-n_F(E+\omega)
\Big)\text{Im}f_1(E+i\eta)~\text{Im}f_2(E+\omega +i\eta),
\label{Seq:Contour_integral}
\eea
where we used $2i\text{Im}f_{1,2}(E+i\eta)=f_{1,2}(E+i\eta)-f_{1,2}(E-i\eta)$.
For large $\omega$,
\bea
&&
P_{ij}(i\omega_m \rightarrow \omega+i\eta)
\approx 
-\frac{4e^2}{\omega^2} T\sum_{iq_n}\int_{\tbf{q}}v_{\tbf{q},i}v_{\tbf{q},j}
\frac{\tilde{z}\tilde{z}_++\xi_\tbf{q}^2+\tilde{\Delta}\tilde{\Delta}_+}{\tilde{z}^2-\xi_{\tbf{q}}^2-\tilde{\Delta}^2}.
\eea
At small $n_{\text{imp}}u_0^2$, the corrections in Eq. \ref{Seq:renormalized} are also small ($\tilde{z}\approx z$, $\tilde{\Delta} \approx \Delta$) and the disorder-mediated response at zero temperature can be obtained by using Eq. \ref{Seq:Contour_integral},
\bea
&&
\text{Re}~\sigma_{\text{dis},ij}(\omega)
=
\frac{\text{Im}P_{ij}(\omega+i\eta)}{\omega}
\approx
\frac{e^2n_{\text{imp}}u_0^2\pi}{\omega^3}\int_0^{\omega}dE
\Big(\int_{\tbf{q}}
v_{\tbf{q},i}
v_{\tbf{q},j}
A(\tbf{q},E)
\Big)
\mathcal{D}(\omega -E)
\Big(1-\frac{\Delta^2}{E(\omega-E)}\Big)
,
\label{Seq:sigma_dis}
\eea
where $A(\tbf{q},\omega)=-\frac{1}{\pi}\text{Tr}\text{Im}G_{\omega+i\eta,\tbf{q}}$ is the spectral function and $\mathcal{D}(\omega)=\int_{\tbf{q}}A(\tbf{q},\omega)$ is the density of state of superconductor. Even if the supercurrent is turned off $\tbf{Q}=0$, Eq. \ref{Seq:sigma_dis} is still finite, $\text{Re}~\sigma_{\text{dis},ij}(\omega)>0$ since the impurity potential Eq. \ref{Seq:H_imp} allows the indirect transitions. For the finite $\tbf{Q}$, the frequency threshold of $A(\tbf{q},\omega)$ is slightly reduced (Fig. 1b in the main text), thus 
$\text{Re}~\sigma_{\text{dis},ij}(\omega)$ is also finite just below the gap edge, $\omega \lesssim 2|\Delta|$.

\end{document}